\documentclass[12pt]{article}
\usepackage{calc}
\usepackage{fancyheadings}
\usepackage{epsfig}
\usepackage{psfrag}
\usepackage{latexsym}
\usepackage{wasysym}
\usepackage{amsfonts}
\usepackage[intlimits]{amsmath}
\usepackage{amsthm}
\usepackage{amssymb}
\usepackage{ifthen}

\newcommand{\field}[1]{\mathbb{#1}}
\newcommand{\N}{\field{N}}
\newcommand{\R}{\field{R}}

\newcommand{\Hi}{{\cal H}}
\newcommand{\De}{{\cal D}}
\newcommand{\Ce}{{\cal C}}
\newcommand{\We}{{\cal W}}
\newcommand{\Be}{{\cal B}}
\newcommand{\tR}{\tilde R}
\newcommand{\tPi}{\tilde \Pi}
\newcommand{\e}{{\rm e}}

\newcommand{\So}[1]{{\mathsf H}_{#1}}
\newcommand{\restrict}{\upharpoonright}

\newtheorem{theorem}{Theorem}
\newtheorem{lemma}{Lemma}
\newtheorem{prop}{Proposition}
\theoremstyle{remark}

\let\norm=\enVert

\newcommand{\nnorm}[1]{\norm{\smash{#1}}}
\newcommand{\abs}[1]{\left\lvert#1\right\rvert}
\newcommand{\nabs}[1]{\abs{\smash{#1}}}
\newcommand{\de}[1]{{\rm d}#1}

\newcommand{\eval}[2][\right]{\relax\ifx#1\right\relax \left.\fi#2#1\rvert}
\newcommand{\inter}[2]{\left[#1,#2\right]}

\DeclareMathOperator{\grad}{grad}
\DeclareMathOperator{\dist}{dist}
\DeclareMathOperator{\supp}{supp}

\DeclareMathOperator{\curl}{curl}

\newenvironment{Cases}{%
  \left\{%
  \array{@{}l@{\quad}l@{}}%
}{%
  \endarray\right.%
}

\makeatletter
\long\def\@makecaption#1#2{%
  \vskip\abovecaptionskip
  \sbox\@tempboxa{{\bf #1:} #2}%
  \ifdim \wd\@tempboxa >\hsize
    {\bf #1:} #2\par
  \else
    \global \@minipagefalse
    \hb@xt@\hsize{\hfil\box\@tempboxa\hfil}%
  \fi
  \vskip\belowcaptionskip}
\makeatother

\begin{document}
\author{J.~Fr\"ohlich \quad G.~M.~Graf \quad J.~Walcher \\ Institute for
  theoretical Physics \\ ETH-H\"onggerberg \\ CH-8093 Z\"urich, Switzerland} 
\date{March 8, 1999}
\title{On the extended nature of edge states of Quantum Hall Hamiltonians}

\maketitle

\begin{abstract}
\noindent
Properties of eigenstates of one-particle Quantum Hall Hamiltonians
localized near the boundary of a two-dimensional electron gas -
so-called edge states - are studied. For finite samples it is shown that edge 
states with energy in an appropriate range between Landau levels remain
extended along the boundary in the presence of a small amount of disorder, in
the sense that they carry a non-zero chiral edge current. 
For a two-dimensional
electron gas confined to a half-plane, the Mourre theory of positive
commutators is applied to prove absolute continuity of the energy spectrum
well in between Landau levels, corresponding to edge states.
\end{abstract}

\section{Introduction and summary of results}
\label{sec:introduction}
In this paper, we study two-dimensional electron gases in a uniform magnetic
field perpendicular to the plane, in the presence of a small amount of
disorder. 
The integer quantum Hall effect, 
discovered by von Klitzing \cite{vKlitzing}, is the
phenomenon that when the Fermi energy of the electron gas is well in between
two Landau levels, the Hall conductivity is equal to an integer multiple of
$e^2/h$. 

Under the
assumption of negligibly small electron-electron interactions, the integer
quantum Hall effect can be derived from a simple one-electron picture. For an
appropriate choice of sample geometry, described by a potential confining the
electrons to the sample, and for a small amount of disorder, one can analyze,
qualitatively, the energy spectrum of the corresponding one-particle
Hamiltonian. In particular, as we show in this paper, eigenenergies
well in between Landau levels correspond to eigenstates localized near, but
extended along, the boundary of the sample, so called edge states. Those edge
states carry a non-zero chiral edge current.
Given a small voltage drop between two parallel components of the boundary, 
the edge states corresponding to the two boundary components will be filled
somewhat asymmetrically with electrons. 
The result is a net Hall current parallel to the boundary and
proportional to the voltage drop. The proportionality factor is the Hall
conductivity. If the Fermi energy of the electron gas is well in between
two Landau levels, and if the voltage drop is small compared to the energy gap
between two adjacent Landau levels and to the Zeeman energy of the magnetic
moment of an electron, the spectral properties of the Hamiltonian yield a Hall
conductivity equal to $e^2/h$ times the number of Landau levels below the
Fermi energy, which is an integer. An argument of this sort, based on a clever
use of gauge invariance, was first given by Laughlin \cite{laughlin} and
subsequently refined by many other people (see e.g. \cite{halperin},
\cite{AS}). The idea that the Hall current is supported by edge states
first appeared in a paper of Halperin \cite{halperin}. The fundamental role of
edge currents in the integer {\em and} the fractional quantum Hall effect was
later understood in terms of a gauge anomaly cancellation mechanism in
\cite{FK} and \cite{wen}.
In this paper, we provide a rigorous analysis of one important detail 
underlying Halperin's argument, namely of the question whether, and in what
sense, the edge states are indeed extended states.

Because we neglect electron-electron interactions, the magnetic
moment of the electron turns out to be essentially irrelevant in our analysis,
and we thus neglect electron spin. 
The one-electron Hamiltonian is therefore given by
\begin{equation}
H= \frac{1}{2m}(\vec{p}-e\vec{A})^{2} + V \text{.}
\label{eq:ham}
\end{equation}
In \eqref{eq:ham}, $m$ is the mass of an electron, $e$ is its charge,
$\vec{A}$ is an electromagnetic vector potential corresponding to a constant
magnetic field $\vec{B}=\curl\vec{A}$, and $V= V_{0} + gV_{d}$ is an external
potential consisting of an edge 
potential, $V_{0}$, that confines the electron to the
sample, and a disorder potential, $gV_{d}$, corresponding to the presence of
random impurities. The factor $g$, a ``coupling constant'',
is a measure for the strength of the disorder. The potential $V_{0}$
can be replaced by appropriate boundary conditions in the definition of the
covariant Laplacian, $(\vec{p}-e\vec{A})^2$, which prevent an electron from
leaving the sample; see, for example, \cite{AANS}, and section \ref{sec:diric}
of the present paper.

\begin{figure}[ht]
  \begin{center}
    \psfrag{B}{\large $\frac{1}{2}\hbar\omega_c$}
    \psfrag{3B}{\large $\frac{3}{2}\hbar\omega_c$}
    \psfrag{5B}{\large $\frac{5}{2}\hbar\omega_c$}
    \psfrag{7B}{\large $\frac{7}{2}\hbar\omega_c$}
    \psfrag{E}{\large $E$}
    \psfrag{Emin}{\large $E_{min}$}
    \psfrag{ran}{edge spectrum}
    \psfrag{lan}{Landau levels}
    \psfrag{hau}{bulk spectrum}
    \includegraphics{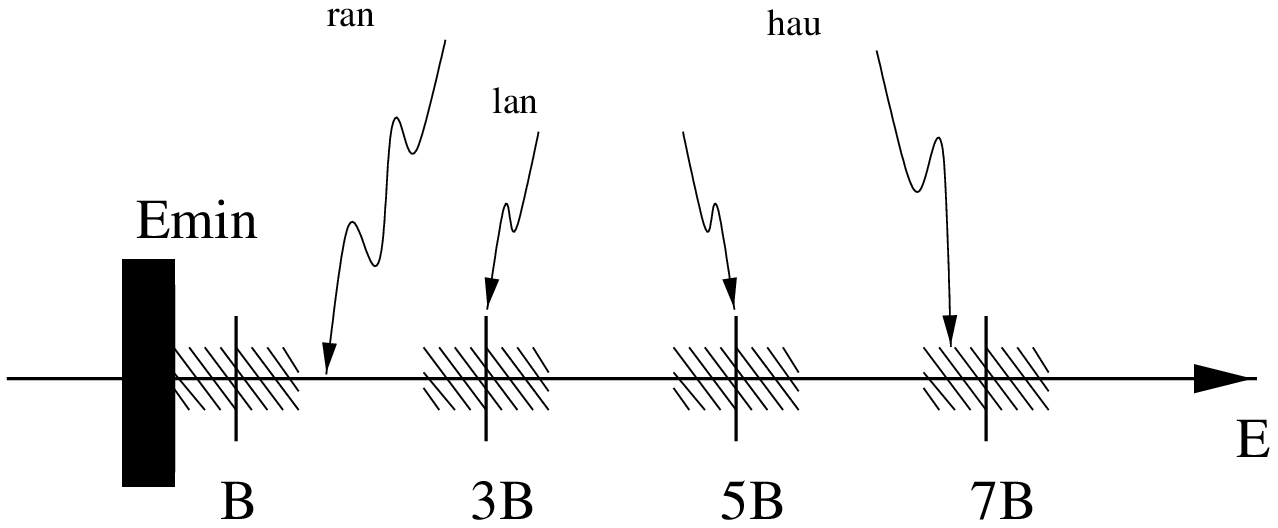}
    \caption{Bulk and edge spectrum of $H$. The energy scale is the
      cyclotron frequency $\hbar\omega_c=eB/m$.}
    \label{fig:spectrum}
  \end{center}
\end{figure}

The location of the energy spectrum of the one-particle Hamiltonian
\eqref{eq:ham} is
indicated in figure \ref{fig:spectrum}. This spectrum consists of a part
corresponding to ``bulk states'' and a part corresponding to ``edge
states''. The former is located near the Landau levels, which are
broadened by the disorder potential. Most of the bulk states are localized,
but close to each Landau level, there are eigenvalues corresponding to
extended bulk states. It is well known that in order to {\em observe} quantum
Hall {\em plateaux}, one needs to have localized bulk states.
The energy spectrum corresponding to edge states is
located in the intervals between the broadened Landau levels. For a sample
covering the entire plane, the intervals between the broadened Landau levels
would be spectral gaps.

The edge states of clean samples ($g=0$) are well understood. For a 
bounded sample and weak disorder, one may 
use analytic perturbation theory in the disorder potential, $gV_d$, in order
to analyze the edge states. Unfortunately, as the sample size increases, the
spacing between eigenvalues of $H$ corresponding to edge states becomes
smaller and smaller, and, as a consequence, the convergence radius of the
perturbation series in $g$ becomes smaller and smaller. Perturbation
theory cannot be used in the limit of an infinitely large sample.

Quantization of the Hall conductivity for small finite samples and weak
disorder is reviewed in section \ref{sec:small}. In our arguments, the
characteristic property of edge states needed to establish the integer quantum
Hall effect is that they carry a non-zero chiral edge current.
This property suggests to {\em define extended states} for arbitrarily large,
bounded samples as states carrying a non-vanishing chiral current. 
In sections \ref{sec:big} and \ref{sec:half}, the extended nature of
edge states is established on the basis of arguments which are valid for
sufficiently weak disorder, i.e. for $\abs{g}<g_*$, but our bounds on $g_*$
are {\em uniform} in the sample size. In sections \ref{sec:small} and 
\ref{sec:big}, our sample has
the original Laughlin cylinder geometry, but our proofs can be adapted 
for other sample shapes, such as the Corbino disc geometry used by Halperin
\cite{halperin}, which we treat in an appendix.

For infinite samples, the natural definition of ``extended states'' is that
they correspond to absolutely continuous spectrum. In section \ref{sec:half},
we 
consider the case of a two-dimensional electron gas confined to a half-plane, 
and prove that the energy spectrum well in between Landau levels is absolutely
continuous for weak disorder. We first model the edge with a smooth but steep
edge potential, and discuss various possible disorder potentials.
It turns out that our bound on the allowed strength of the 
disorder becomes smaller as the edge is made steeper. We then treat an
``infinitely steep'' edge directly by introducing Dirichlet boundary
conditions in the definition of the  covariant Laplacian. We show that for
weak disorder, the edge states are again extended states. The proofs in
section \ref{sec:half} are based on an application of the
Mourre theory of positive commutators \cite{mourre}, which is briefly
presented in section \ref{sec:methods}. In section \ref{sec:methods}, 
we also explain more specifically our assumptions about the disorder and edge
potentials. 
 
For both finite and infinite samples, the extended nature of the edge states
is analyzed with the help of the so-called ``guiding center'' of cyclotron
motion. The commutator of the coordinate of the guiding center along the edge
with the Hamiltonian is given by the derivative of the potential in the
direction perpendicular to the edge\footnote{This is in the case of an edge
potential, for Dirichlet boundary conditions, see section \ref{sec:diric}}.
The proofs are reduced to showing that this commutator is positive on states
with energy well in between Landau levels. Instead of the guiding center, one
can also use the coordinate of the particle itself along the edge as conjugate
operator in the sense of Mourre theory. For the problem with Dirichlet boundary
conditions, De Bi\`evre and Pul\'e \cite{DBP} have shown that this allows to
relax the assumptions on the disorder potential. It turns out that a Mourre
estimate for one commutator is equivalent to a Mourre estimate for the other
with the same lower bound on the commutator, but the techniques used to prove
the estimate are different. 

Recently, Macris, Martin, and Pul\'e (see \cite{MMP}) have studied 
the half-plane case by a somewhat different method. They rule out the
existence of eigenvalues between the broadened Landau levels by showing that
the expectation value of the derivative of the potential in the direction
perpendicular to the edge would be positive in an assumed eigenstate with
energy between the broadened Landau levels. This would contradict the fact
that the expectation value of a commutator with the Hamiltonian in an energy
eigenstate must vanish by the virial theorem. To prove the positivity of the
commutator in an assumed eigenstate, for weak disorder, they estimate the
decay of edge state eigenfunctions into the edge with the help of Brownian
motion techniques. Our use of the conjugate operator method allows us to
exclude not only point spectrum, but also singular continuous
spectrum. Furthermore, the methods we use to estimate the commutator
allow a clear discussion of the assumptions about the disorder potential,
and we also treat the problem with Dirichlet boundary conditions, whereas the
estimates for smooth potentials tend to fail in the limit of an infinitely
steep edge.


\section{The Laughlin argument revisited}
\label{sec:small}
\begin{figure}[ht]
  \begin{center}  
    \psfrag{j}{${I}_{\varphi}$}
    \psfrag{VH}{$V_H$}
    \psfrag{L}{$L$}
    \psfrag{R}{$R$}
    \psfrag{Phi}{$\Phi$}
    \psfrag{phi}{$\phi$}
    \psfrag{B}{$B$}
    \psfrag{y}{$y$}
    \includegraphics{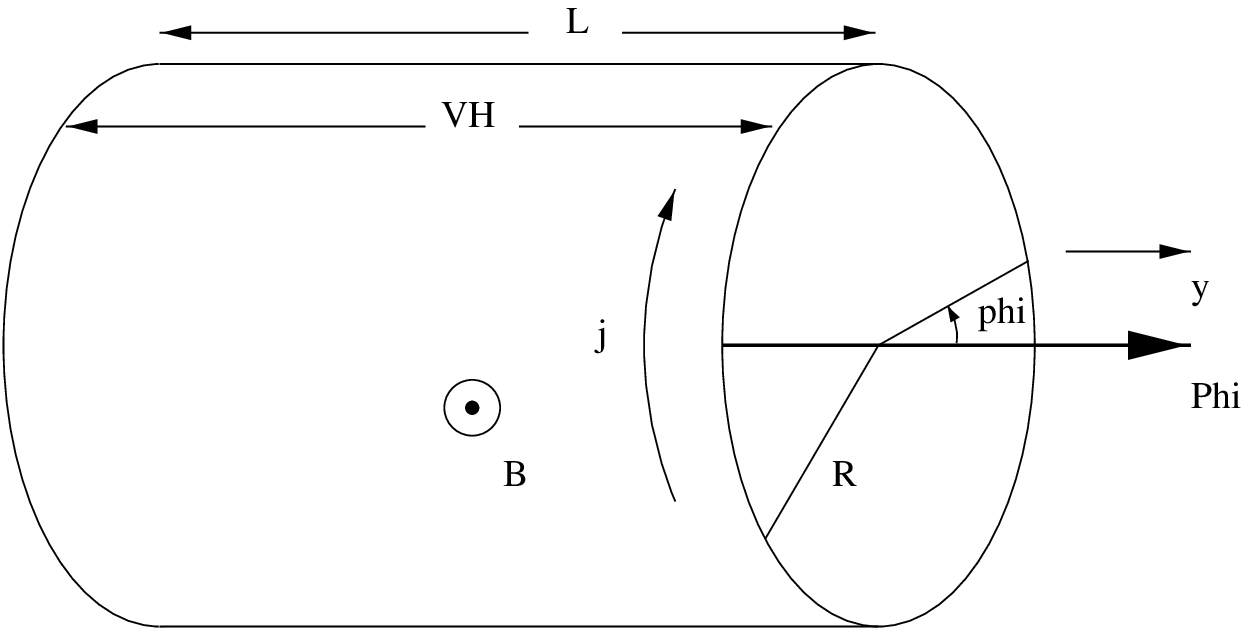}
    \caption{Cylinder geometry}
    \label{fig:cylinder}
  \end{center}
\end{figure}

In this section, we review the argument leading to the integer quantization of
the Hall conductivity, motivating our interest for the extended nature of edge
states. In order to keep our analysis as simple as possible, we consider the
cylinder geometry used in the original Laughlin argument \cite{laughlin}. The
Hall current flows along the circumference and the Hall voltage is measured
between two edge circles (see figure \ref{fig:cylinder}). In addition to the
homogeneous magnetic field perpendicular to the surface of the cylinder, there
is a ``magnetic flux tube'', $\Phi$, at the axis of the cylinder. 

The cylinder is characterized by two length scales, the radius, $R$, and the 
distance, $L$, between the two edges. Both lengths play a role in the
mathematical analysis. Increasing $R$ reduces the spacing between edge state
eigenvalues of the Hamiltonian, and thus limits the applicability of
perturbation theory to analyze the edge states. On the other hand, $L$
influences the tunneling probability between two edges. Physically, we expect
that for weak disorder, the tunneling probability per edge length for states
with energy well in between Landau levels is suppressed exponentially in
$L/l_c$, where $l_c$ is the cyclotron length. In this paper, we shall not
provide rigorous bounds on those tunneling rates, but only deal with the
problem connected to the spacing of eigenvalues. In the proofs in sections
\ref{sec:big} and \ref{sec:half}, possible tunneling between edges will be
avoided by considering only one edge, or, equivalently, taking one edge to
infinity. In the present section, we still consider two edges.

In the Corbino disc or annulus geometry introduced by Halperin
\cite{halperin}, one cannot completely eliminate the tunneling problem,
because, even if one considers only the outer edge of the annulus, the flux
tube at the center is comparable to having a second edge, in the sense that
for generic $\Phi$, there are eigenvalues between Landau levels. Without
precise estimates on tunneling probabilities between inner and outer edge, we
can only show that edge states are extended for small $\abs{\Phi}$. The
argument for the Corbino disc is carried out in appendix \ref{sec:app1}.

The coordinate along the axis of the cylinder will be denoted by $y$, and the
coordinate perpendicular to it will be $x=R\varphi$, where
$0<\varphi\le2\pi$. The magnetic field is pointing radially outward, and the
vector potential is chosen in $\varphi$-direction 
$A_{\varphi}= -By + \Phi/2\pi R$. Of course, the magnetic field can only
be homogeneous on the two-dimensional surface of the cylinder, since otherwise
the Maxwell equations would be violated.
In these coordinates, the Hamiltonian is 
\begin{equation}
H= \frac{1}{2m}\left(-\partial_y^2  +
  \left(\frac{1}{iR}\partial_\varphi
-e\left(-By+\frac{\Phi}{2\pi R}\right)\right)^2\right) +
V(y,\varphi),
\label{eq:hamcyl}
\end{equation}
in units where $\hbar=1$. We start with $V=0$, that is, with a cylinder
infinite in $y$-direction and without disorder. The states can be labeled
by the angular momentum quantum number $l$ and the Landau band index $n$. 
The energy depends only on $n$ through $E_{n,l}=(n+1/2)\omega_c$. In the
$y$-direction, the eigenfunctions are harmonic oscillator wave functions, 
localized near $y_0(l-e\Phi/2\pi)=(-l+e\Phi/2\pi)/eBR$.
Changing $\Phi$ does not affect the energy of the states, but only their
position along the cylinder. In particular, a change of $\Phi$ by $2\pi/e$ maps
states with angular momentum $l$ on those with angular momentum $l-1$.
This ``spectral flow'' produced by the change in $\Phi$ plays an important
role in the following arguments.  

For a symmetric confining potential $V_0=V_0(y)$, it is 
possible to continue labeling the states by $l$ and $n$
and to qualitatively discuss the dependence of the energy $E_{n,l}(\Phi)$ on
$l$ and $\Phi$. The one-dimensional Hamiltonian 
for the motion in $y$-direction that results after separating the angular
momentum, is analytic in the parameter $l-e\Phi/2\pi$.
Therefore, $E_{n,l}(\Phi)=E_n(l-e\Phi/2\pi)$ are analytic functions, and the
spectral flow of eigenstates with changing $\Phi$ is preserved by a symmetric
$V_0$, 
\begin{equation}
E_{n,l}(\Phi+2\pi/e)=E_{n,l-1}(\Phi).
\label{eq:flow}
\end{equation}
Furthermore, all eigenstates are well localized in the $y$-direction and the
localization position, $y_0(l-e\Phi/2\pi)$, is also an analytic function, which
is monotonically decreasing as can be seen by inspection of the Hamiltonian
\eqref{eq:hamcyl}. $E_{n,l}(\Phi)$ can lie between Landau levels only if
$y_0(l-e\Phi/2\pi)$ comes close to an edge of the cylinder, that is, only for
edge states. For each $n$, it is possible to identify those $l$ which
correspond to states at the left and right edge. Large positive $l$ correspond
to the left edge, and large negative $l$ to the right edge.

Consider the current carried by an $(n,l)$-state in $\varphi$-direction, 
\begin{equation}
I_{\varphi,n,l} = -\frac{d E_{n,l}(\Phi)}{d\Phi}.
\end{equation}
Under the assumption that $V_0$ is monotonically increasing as one leaves the 
sample on either edge of the cylinder, so that it correctly describes the
confining of the electron gas to the sample, it is easy to see that
$I_{\varphi,n,l}$ has a definite sign for states localized at either
edge. Edge states carry a chiral edge current.

Before considering the influence of disorder on the spectral properties of
$H$, we show that the spectral flow \eqref{eq:flow} implies the integer
quantum Hall effect.

Assume that the Fermi energy, $E_F$, of the electron gas on the surface of the
cylinder lies between two Landau levels, and that there is a small voltage
drop, $eV_H=\mu_r-\mu_l>0$, between the left and right edge. Let $\nu$ be the
number of Landau levels below $E_F$.  
For each $n\le\nu$, there is an $l_{min}(n)$ with
$E_{n,l}(0)>\mu_r$ for all $l\le l_{min}(n)$ and an $l_{max}(n)$ with
$E_{n,l}(0)>\mu_l$ 
for all $l>l_{max}(n)$. We shall use that, approximately,
$E_{n,l_{max}(n)}(0)=\mu_l$ and $E_{n,l_{min}(n)}(0)=\mu_r$. The total current
carried by the electron gas is 
\begin{equation}
I_{\varphi}= \sum_{n=0}^{\nu} \;\; \sum_{l=l_{min}(n)+1}^{l_{max}(n)}
    -\frac{d E_{n,l}(\Phi)}{d\Phi}.
\label{eq:summe}
\end{equation}
Note that we here sum over all states below the Fermi energy, not only the
edge states. Since the flux quantum $2\pi/e$ is a small quantity, and if the
number of contributing states is large, the current will be approximately the
same as its $\Phi$-average over a range of $2\pi/e$. 
\begin{align}
I_{\varphi} &= \sum -\frac{e}{2\pi}\int_0^{2\pi/e} \de{\Phi} 
\frac{d E_{n,l}(\Phi)}{d\Phi}, \\
\intertext{which, using \eqref{eq:flow}, becomes}
&= \sum_{n=0}^{\nu} -\frac{e}{2\pi}\left(E_{n,l_{min}(n)}(0) -
  E_{n,l_{max}(n)}(0)\right) \\ &= -\frac{e^2}{2\pi} \nu V_H.
\end{align}
The Hall conductivity $\sigma_H= \abs{I_{\varphi}/V_H}$ is thus an integer
multiple of $e^2/2\pi=e^2/h$.

\begin{figure}[ht]
  \begin{center}
    \psfrag{EF}{$E_F$}
    \psfrag{mul}{$\mu_l$}
    \psfrag{mur}{$\mu_r$}
    \psfrag{l}{$-(l-e\Phi/2\pi)$}
    \psfrag{E}{$E$}
    \includegraphics{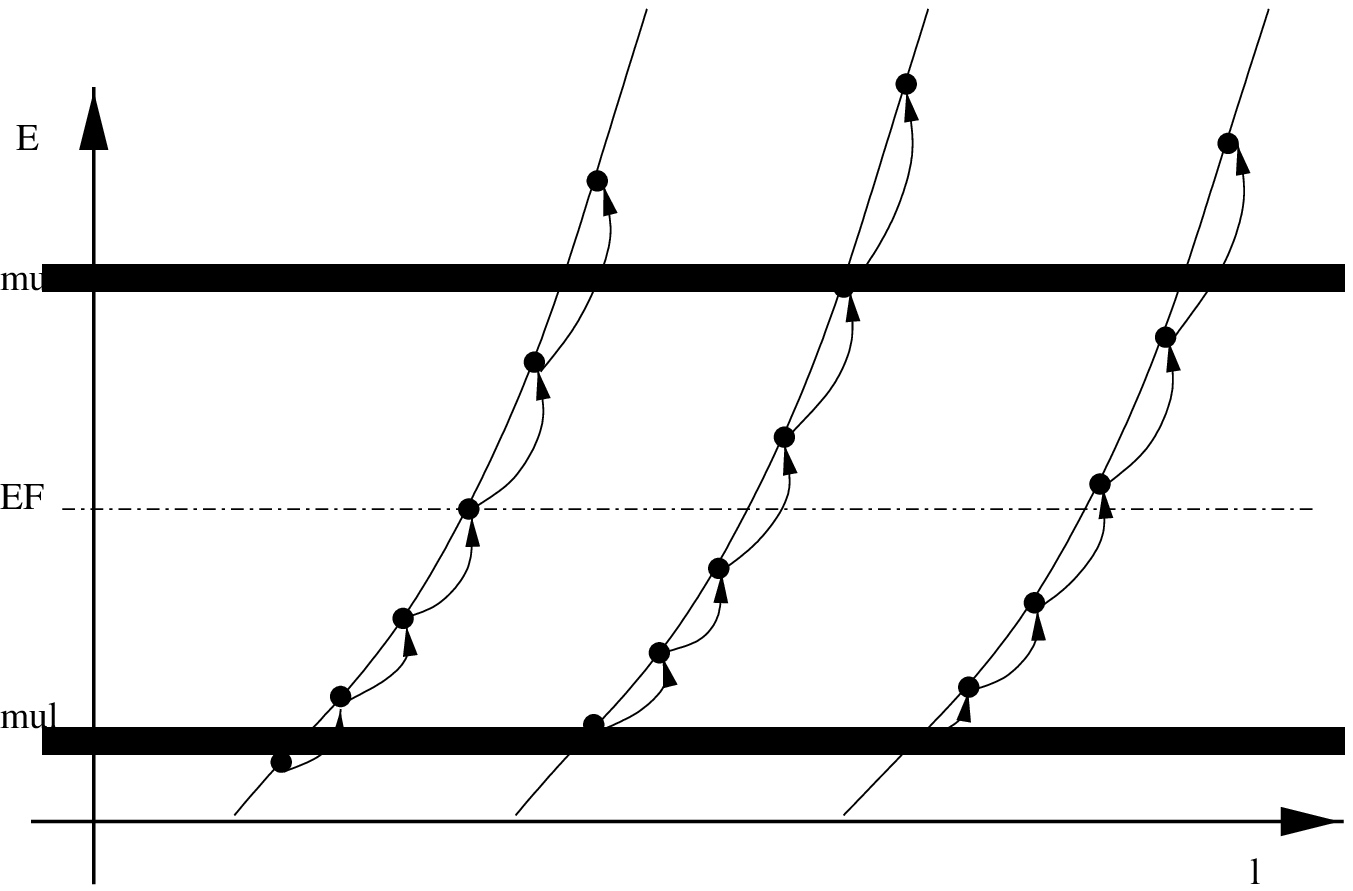}
    \caption{The spectral flow of edge states under variation of $\Phi$. Each
      line represents a Landau band with eigenvalues symbolized by a
      dot. $\mu_l$ and $\mu_r$ are the chemical potentials on the left and
      right edge of the cylinder, respectively.}
    \label{fig:specflow}
  \end{center}
\end{figure}

It is clear that in this derivation, the Hall current is due to the chiral
currents carried by the edge states. The sum in \eqref{eq:summe} could also be
performed only over those states with energy between $\mu_l$ and $\mu_r$,
localized near one edge of the cylinder. The 
approximations we have used above, namely the averaging of the current over
$\Phi$ from $0$ to $2\pi/e$ and the substitution
$E_{n,l_{max/min}(n)}=\mu_{l/r}$, become better as the spacing between edge
state eigenvalues decreases with increasing sample size, $R$. 
The necessary finite size corrections can in principle be estimated, but we
omit this here.

We further stress that the possibility to label the states by angular
momentum, $l$, and Landau band index, $n$, is not crucial for
$\sigma_H/(e^2/h)$ to be an integer.  
The important point of the derivation is that a change of
$\Phi$ by $2\pi/e$ moves a definite number of edge states through the interval
$[\mu_l,\mu_r]$. To see this more clearly, assume, for a more general
situation, a labeling of states by an
index $\alpha$ instead of $n$ and $l$, with energies $E_{\alpha}(\Phi)$.
Because the spectrum of $H$ is strictly invariant under a change of $\Phi$ by
$2\pi/e$, there is a bijective map $\alpha\mapsto\beta(\alpha)$ with
$E_{\alpha}(2\pi/e)=E_{\beta(\alpha)}(0)$. Denote by ${\cal I}$ those $\alpha$
with $\mu_l\le E_{\alpha}(0)\le\mu_r$. Up to finite size corrections, the Hall
current is given by 
\begin{equation}
I_{\varphi}= - \frac{e}{2\pi} \Bigl(\sum_{\beta({\cal I}) \setminus {\cal I}}
E_{\alpha}(0) -\sum_{{\cal I} \setminus \beta({\cal I})} E_{\alpha}(0)\Bigr).
\end{equation}
We claim that the central property of edge states is that they satisfy an
estimate of the form
\begin{equation}
\frac{C}{R} \ge \frac{d{E_{\alpha}(\Phi)}}{d{\Phi}} \ge \frac{C'}{R}.
\label{eq:haw}
\end{equation}
where $C$ and $C'$ are nonzero constants whose {\em common sign} depends upon
whether the edge states are localized on the right or on the left edge, and
$R$ is the size of the sample. 
The estimate \eqref{eq:haw} is satisfied in the situation of a clean sample
$(g=0)$ described above. If we assume it in the general situation, it
implies that $C/R \ge \bigl(E_{\beta(\alpha)}(0)-E_{\alpha}(0)\bigr) e/2\pi \ge
C'/R > 0$, if $\alpha$ corresponds to an edge state at the right
edge. Therefore,  
\begin{equation}
E_{\alpha}(0) =
\begin{cases} \mu_r +O(1/R) & \text{if $\alpha \in \beta({\cal I})\setminus
    {\cal I}$} \\
\mu_l + O(1/R) & \text{if $\alpha\in {\cal I}\setminus\beta({\cal I})$,} 
\end{cases}
\end{equation}
Because $\beta$ is bijective,
$\abs{{\cal I}\setminus\beta({\cal I})} = \abs{\beta({\cal I})\setminus {\cal
    I}}$, and we see that
\begin{equation}
\frac{\sigma_H}{(e^2/h)} = \abs{\beta({\cal I})\setminus {\cal I}}
\left(1+O(1/R)\right) 
\end{equation} 
is an integer, up to finite size corrections.

In this form, the argument completely clarifies the universal character of the
integer quantum Hall effect. The effect is not related to any particular
geometry or symmetry of the sample.
The only necessary ingredients are that states with
energy between Landau levels are localized near the edges of the sample and
that those edge states are extended states, in the sense that they carry a
chiral edge current, or, in other words, that they satisfy an estimate of the
form \eqref{eq:haw}. The definition of extended states as states whose
energy changes when $\Phi$ is varied is consistent with the fact that 
an eigenfunction of $H$ whose support does not surround the flux will up to
a phase factor not be affected by a change of $\Phi$, since, in a simply
connected region, the influence of $\Phi$ can always be gauged away.

We now return to the more special situation of the cylinder geometry
and include the disorder potential, $gV_d$, in our discussion. By analytic
perturbation theory, we know that the eigenenergies, $E_{\alpha}(\Phi,g)$,
are analytic functions in $g$. For a generic edge potential $V_0$, and at some
generic value of the flux, $\Phi_0$, eigenenergies which lie between Landau 
levels will be non-degenerate. As long as $gV_d$ is small as
compared to the spacing between the energy levels at $g=0$, the labels $n$ and
$l$ are well defined for this $\Phi_0$. By appropriately identifying the
$\alpha$'s, this defines the functions $E_{n,l}(\Phi,g)$ for energies between
Landau levels, for all $\Phi$, although, for particular values of $\Phi$,
energies of states may become degenerate. By construction, the energies
$E_{n,l}(\Phi_0,0)$ agree with the previously defined $E_{n,l}(\Phi_0)$. We
assume that this is true for all values of $\Phi$. This assumption can only be
justified by precise estimates on tunneling probabilities between edges.
Because the eigenfunctions are also analytic in $g$, labels $(n,l)$ that
correspond to edge states for $g=0$ will also correspond to states localized
at the edge for $g$ non-zero, but small. Therefore, if we can show the analog
of equation \eqref{eq:flow}, 
\begin{equation}
E_{n,l}(\Phi+2\pi/e,g)=E_{n,l-1}(\Phi,g),
\label{eq:flowd}
\end{equation}
this will imply the integer quantum Hall effect. 

The shift in energy due to the disorder, $E_{n,l}(\Phi,g)-E_{n,l}(\Phi,0)$, is
of the order of $g$ by the Feynman-Hellmann theorem, and \eqref{eq:flow} then
immediately implies that $E_{n,l}(\Phi+2\pi/e,g)-E_{n,l-1}(\Phi,g)$ is also of
the order of $g$. On the other hand, a change of
$\Phi$ by $2\pi/e$ can be compensated by a gauge transformation and does not
change the spectrum at all. Therefore, there must be an $n'$ and an $l'$ with
$E_{n,l}(\Phi+2\pi/e,g)=E_{n',l'}(\Phi,g)$. But if, without
disorder, the energies are sufficiently far apart and non-degenerate, for
weak disorder, this is only possible for $n'=n$ and $l'=l-1$. This implies 
\eqref{eq:flowd}.

The perturbative argument for the spectral flow obviously breaks down when the
radius of the cylinder becomes very large, because the spacing between
eigenvalues at $g=0$ decreases with increasing $R$. On the other hand, as we
have already noted, finite size corrections become small at least inversely
proportional to $R$. It is therefore desirable to have an argument that
establishes the spectral flow with a bound on the allowed strength of the
disorder that is {\em uniform} in $R$. Such an argument will be provided in
section \ref{sec:big}. The rigorous proofs establishing the bounds in
\eqref{eq:haw} are contained in section \ref{sec:edge}.


\section{Methods and tools}
\label{sec:methods}
\subsection{The guiding center}
\label{sec:conjop}

It is well known that the classical cyclotron orbit of a charged particle in a
homogeneous magnetic field drifts under the influence of an electrostatic
potential. This can be seen most simply by considering the ``guiding center''
of the motion, which is the center of a circle with cyclotron radius
$r=v/\omega_c= v/(eB/m)$ that passes through the position of the particle. The
velocity,$v$, of the particle is given by 
\begin{equation}
\vec{v}=\frac{1}{m}(\vec{p}-e\vec{A}).
\end{equation}
This yields for the guiding center
\begin{equation}
\vec{Z} = \vec{r} + (\vec{p}-e\vec{A})\times \frac{\hat{B}}{B},
\end{equation}
where $\hat B$ is a unit vector in $B$-direction perpendicular to the plane. 


It is easily established that the equation of motion of $\vec{Z}$ in a
potential $V$ is
\begin{equation}
\dot{\vec{Z}}= \frac{\hat B}{B} \times \grad V.
\end{equation}

The separation of the motion of the guiding center from the cyclotronic motion 
will be used in section \ref{sec:big} to get an estimable expression for the
azimuthal current carried by any eigenstate of the Hamiltonian. 
In section \ref{sec:half}, we will use the coordinate of the guiding center 
along the edge as a conjugate operator in the sense of Mourre theory. 

\subsection{Positive commutators and absolutely continuous spectrum}
\label{sec:poscomm}

If in a classical Hamiltonian system, one can show that for some orbit the
Poisson bracket of some coordinate $K$ with the Hamiltonian $H$
remains bounded away from zero, $\dot K = \{K,H\} \geq \alpha >0$, for all
times, one can conclude that the motion is extended along this coordinate.
The quantum mechanical counterpart of this simple statement is the
following: Assume there exists a ``conjugate operator'', $A$, such that the
commutator of $A$ with the Hamiltonian is positive on some energy interval
$\Delta$, 
\begin{equation}
E_{\Delta}(H)\left[ H,iA \right] E_{\Delta}(H) \geq \alpha
E_{\Delta}(H),  
\label{eq:mouest} 
\end{equation}
with $\alpha > 0$, and where $E_{\Delta}(H)$ denotes the spectral projector of
$H$ on $\Delta$. Noting that if $\psi$ is an eigenstate of $H$, we have 
$\left(\psi,\left[H,iA\right]\psi\right)=0$ by the virial theorem, we can
conclude from \eqref{eq:mouest}
that $H$ can not have an eigenvalue in the interval $\Delta$. It was
first proved by Mourre \cite{mourre} that under additional regularity
assumptions on $H$ and its commutators with $A$, the spectrum of $H$ is
actually purely absolutely continuous on $\Delta$.
Equation \eqref{eq:mouest} is termed a Mourre estimate. 

The necessary assumptions on $H$ have been subsequently relaxed considerably;
see, for example, \cite{cycon}, \cite{sahbani}, or \cite{ABG}.
For the treatment of the problem with a smooth, steep edge potential in section
\ref{sec:edge}, the original assumptions of Mourre can be verified. Those are
(see \cite{mourre}): 
\begin{enumerate}
\renewcommand{\labelenumi}{(\roman{enumi})}
\renewcommand{\theenumi}{(\roman{enumi})}
\setlength{\parskip}{0cm}
\item \label{pcassum1} $H$ and $A$ are self-adjoint operators with domains
  $\De(H)$ and $\De(A)$. $\De(H)\cap\De(A)$ is a core for $H$.
\item \label{pcassum2} The unitary group $e^{iAa}$ generated by $A$
  leaves $\De(H)$ invariant and for all $\psi\in\De(H)$
\begin{equation*}
\sup_{\abs{a}<1}\norm{He^{iAa}\psi} < \infty.
\end{equation*}
\item \label{pcassum3} The quadratic form $\left[H,iA\right]$ which is defined
  on $\De(H)\cap\De(A)$, is bounded below and closable; the associated
  self-adjoint operator admits a domain containing $\De(H)$.
\item \label{pcassum4} The quadratic form $\bigl[[H,iA],iA\bigr]$
 is form-bounded by $\abs{H}^2$.
\end{enumerate}

Following Mourre's work, the main efforts have gone in the direction of
reducing the regularity assumptions described in \ref{pcassum3} and
\ref{pcassum4}. We do not know of any results if the conjugate operator is not
self-adjoint. In the case of Dirichlet boundary conditions in section
\ref{sec:diric}, we are precisely in such a situation, and we will use several
tricks to fit the problem into the framework treated in \cite{sahbani}. Since
a virial theorem is valid for non self-adjoint operators, too, we expect that
the conjugate operator method might have a generalization to such situations.

\subsection{Decay of edge state eigenfunctions}

The decay of edge state eigenfunctions into the bulk can be proved with the
help of the equation
\begin{equation}
\label{eq:decay}
\psi = (E-H_0-V_d)^{-1} V_0\psi,
\end{equation}
where $\psi$ is an eigenfunction of $H$ with energy $E$ in the gaps of the
bulk spectrum, and $H_0= (\vec{p}-e\vec{A})^2/2m$, $H=H_0+V_0+V_d$.

The free resolvent $(E-H_0)^{-1}$ can be calculated explicitly in coordinate
space representation, and it has Gaussian decay. Using this decay, equation 
\eqref{eq:decay}, and the fact that $V_0\psi$ is supported at the edge, one
can show that the eigenfunction decays exponentially into the bulk of the
sample.

More technical details can be found in appendix \ref{sec:app1}. 

\subsection{Further assumptions}

The disorder potentials under consideration are always bounded and small
compared to the cyclotron energy. In some places, we also use bounds on the
derivatives of the potential, but this is presumably not strictly necessary.

Our smallness assumption on the disorder potential can very likely be 
replaced by a smallness assumption on the variance of a random potential, by
using methods from Anderson localization theory.

As mentioned in section \ref{sec:small}, we will avoid the problems caused by
possible tunneling or resonances between two edges by limiting our analysis to
a single edge. The main point in this paper is to show that edge states are
extended, which is a basic ingredient in a proof of the quantization of the
Hall conductivity. 

The edge potentials that are allowed by our proofs include all kinds of
bounded as well as unbounded, polynomially or exponentially growing
potentials. They are assumed to vanish 
in the bulk, and be monotonically increasing in the edge. 


\section{Extended edge states in large cylindrical samples}
\label{sec:big}
In section \ref{sec:small}, the notion of an ``extended edge state'' for
arbitrarily large, bounded samples was defined as one that carries a chiral
edge current. The central estimate needed for the derivation of the integer
quantum Hall effect is \eqref{eq:haw}. In the present section, we show that if
$\psi$ is an eigenstate of the Hamiltonian $H$, with an energy between Landau
levels, and if the disorder is weak, the current carried by $\psi$ is
non-vanishing and has a definite sign.  

We assume the cylinder geometry described in section \ref{sec:small}, with an
edge potential confining the electron to the region $y<0$. The current in
$\varphi$-direction can be written as 
\begin{equation}
I_{\varphi} =-\frac{\partial E}{\partial \Phi}= \left(\psi,\frac{2}{2\pi
    R}\Bigl( \frac{\partial_{\varphi}}{iR} - \frac{\Phi}{2\pi R} + By\Bigr)
  \psi \right).
\label{eq:curr}
\end{equation}
Here and from now on, units are chosen in which $m=1/2$ and $e=1$.
Because $\psi$ is an eigenstate of $H$, the expectation value of 
\begin{equation}
[H,ip_y]= -B\Bigl(\frac{\partial_{\varphi}}{iR} - 
\frac{\Phi}{2\pi R} + By\Bigr) - \partial_y\!V
\end{equation}
in $\psi$ vanishes by the virial theorem. This yields the equation
\begin{equation}
I_{\varphi} = -\frac{1}{2\pi BR} \bigl(\psi, \partial_y\!V\psi\bigr).
\label{eq:super}
\end{equation}
Equation \eqref{eq:super} can be interpreted as saying that 
the current arises solely from the motion of the guiding center, whereas the
cyclotronic motion does not contribute to the current when averaged over the
whole cylinder. 

It is now intuitively clear why the current is chiral and non-vanishing.
$\bigl(\psi, \partial_y\!V\psi\bigr)$ gets a positive contribution from the
edge potential, and a contribution of indefinite sign from the disorder. If
the disorder is weak, and because an edge state is localized near the edge,
the contribution from the edge potential dominates the one from the disorder
potential, so that the claim follows. The calculations needed to rigorously
establish the lower bound on $\bigl(\psi, \partial_y\!V\psi\bigr)$
can be found in section \ref{sec:edge} (proof of Theorem \ref{thm:mourreest1}).  

On the other hand $(\partial_{\varphi}/iR-\Phi/2\pi R+By)\psi$ is
bounded in norm by the Hamiltonian. Equations \eqref{eq:curr} and
\eqref{eq:super} therefore imply the estimate \eqref{eq:haw}.


\section{The half-plane}
\label{sec:half}
In this section, we consider the half-plane geometry. The magnetic field $B$
points in the $z$-direction, the sample is infinite in the $x$-direction, and
the electron gas is confined by a wall to the region $y<0$. 

The half-plane can be viewed as the limiting case $R\rightarrow\infty$ of the
cylinder geometry with one edge. The spectrum of the Hamiltonian is not purely
discrete anymore and it is not possible to induce a spectral flow by
changing a flux $\Phi$. The definition of extended states is that the
corresponding spectrum is absolutely continuous.

\subsection{The steep edge}
\label{sec:edge}
We first treat the case of a smooth edge potential, $V_0$, which
vanishes for $y<0$ and is rapidly increasing for $y>0$. The total potential is
$V=V_0+V_d$, and the Hamiltonian is
\begin{equation} 
H= (\vec{p}-\vec{A})^2 + V =H_0+V=H_0+V_0+V_d,
\end{equation}
where we have again chosen units with $e=1$ and $m=1/2$. The vector potential
is taken in Landau gauge $A_x = -By$, $A_y =A_z=0$. We show absolute
continuity for parts of the spectrum of $H$ located between Landau levels,
using Mourre theory with the $x$-coordinate of the guiding center as conjugate
operator.

The case $V=V_0$, that is without disorder potential, is standard.
In Landau-gauge, the $y$-coordinate of the guiding center $Z_y=-p_x/B$ is a
cyclic coordinate. After a Fourier transformation in the $x$-direction, the
problem splits into one-dimensional Hamiltonians $H_k$ indexed by the
constant of motion $k=-BZ_y$. Those have spectrum $E_n(k)$, where $n$ is
the Landau band index. For $k\rightarrow\infty$, one can easily establish
$E_n(k)\rightarrow (2n+1)B$, while for $k\rightarrow -\infty$, we have 
$E_n(k)\rightarrow\infty$. Also $E_n(k)$ is analytic as a function of
$k$, so this implies that the spectrum of the full Hamiltonian is absolutely
continuous (see \cite{RS}, Theorem XIII.16).

After introducing a disorder potential, it is worthwhile to first estimate the
changes in the location of the spectrum. If the disorder is a random potential
satisfying certain reasonable assumptions, it is known that the almost sure
spectrum of the full Hamiltonian contains the spectrum of the clean
Hamiltonian as a subset. More details about this will be found in appendix
\ref{sec:app2}. The situation is of course more complicated for an arbitrary
deterministic potential, but the results about continuity of the spectrum do
not depend upon existence of spectrum.

In the chosen gauge, our conjugate operator is $\Pi=BZ_x=p_y+Bx$. 
The commutator with $H$ is $[H,i\Pi]=-\partial_y\!V$, so that one
rather has a ``negative commutator'' than a positive one, but this does
obvioulsy not hinder the application of Mourre theory. 

In addition to establishing a Mourre estimate, we
need to assume that the edge potential allows the verification of 
conditions \ref{pcassum1} to \ref{pcassum4} from section
\ref{sec:poscomm}. We shall not make the attempt to present the optimal
conditions on $V_0$, but simply note that \ref{pcassum3} and \ref{pcassum4}
are valid, for example, if the potential is an upper bound for its own
derivatives.  
Assumption \ref{pcassum1} is trivially valid since the $C^{\infty}$
function with compact support, $C^{\infty}_c$, form a core for both $H$ and
$\Pi$. As for \ref{pcassum2}, note that up to a phase factor, 
the group generated by $\Pi$ are the translations in $y$-direction. If the
edge potential $V_0=V_0(y)$ does not increase too fast, for example
subexponentially, so that an estimate of the form $V_0(y+\alpha)\le C V_0(y)$
holds uniformly in $y$, the domain of $V_0$ is invariant under those
translations, and with it also the domain of $H$, since the domain of the
kinetic energy is trivially invariant. 
As noted above, extensions of the Mourre theory allow the treatment of
much more singular potentials.

For the following theorem, we assume an unbounded edge potential, vanishing
for $y<0$, with $V_0'(y)\ge 0$ for all $y$ and $\inf{\{V_0'(y);y\ge b\}}>0$
for all $b>0$. We discuss afterwards how the assumption that $V_0$ is
unbounded can be avoided.

\begin{theorem}[Mourre estimate]
\label{thm:mourreest1}
Assume $E\notin\sigma(H_0)=\{(2n+1)B,n\in\N_0\}$. Then there is a constant
$\delta$, 
such that if the disorder potential satisfies $|V_d|\leq\delta$, there is an
open interval $\Delta\ni E$ and a positive constant $\alpha$ with
\begin{equation}{\label{eq:mourreest1}}
-E_{\Delta}(H)\,[ H,i\Pi] \,E_{\Delta}(H) \geq \alpha E_{\Delta}(H).
\end{equation} 
\end{theorem}

The strategy for the proof is clear: Since $-\left[H,i\Pi\right] =
\partial_y\!V_0+\partial_y\!V_d$, one first establishes the estimate
considering only $\partial_y\!V_0$, but including $V_d$ in the
Hamiltonian. This is the content of proposition \ref{thm:proposition1}, which
yields a bound $E_{\Delta}(H)\partial_y\!V_0E_{\Delta}(H)\ge\tilde\alpha
E_{\Delta}(H)$. Then one estimates $\abs{\partial_y\!V_d}$ on
$\Delta$ by $E$ and $\delta$, and Theorem \ref{thm:mourreest1} follows if
$\left\lvert E_{\Delta}\,\partial_y\!V_d\,E_{\Delta}\right\rvert<\tilde\alpha$.
It is also possible to introduce another constant $\delta'$ to control the
derivative $\abs{\partial_y\!V_d}\le\delta'$ separately, so that
\eqref{eq:mourreest1} follows for $\delta'<\tilde\alpha$. 
This allows a somewhat more generous choice of $\delta$.

We point out again that the proof of Theorem \ref{thm:mourreest1} can be
transfered, without any changes, to the cylinder geometry to prove that
$\bigl(\psi,\partial_yV\psi\bigr)$ is positive if $\psi$ is an energy
eigenstate with energy well in between Landau levels. 

\begin{prop}
\label{thm:proposition1}
Assume $E\notin\sigma(H_0)$. Then there is a constant $\delta$, such
that if the disorder satisfies $|V_d| \leq\delta$, there is an open interval 
$\Delta\ni E$ and a positive constant $\tilde\alpha$ with
\begin{equation}
E_{\Delta}(H)\,\partial_y\!V_0\, E_{\Delta}(H) \geq \tilde\alpha
E_{\Delta}(H).   
\end{equation}
\end{prop}
\newcommand{\Ga}{\bigl(\psi,V_0'\psi\bigr)}

This proposition is at the heart of the matter, and its proof will be
presented in some detail.  

We have to show that $\Ga\ge\tilde\alpha\norm{\psi}^2$ with $\tilde\alpha>0$
holds for all $\psi$ with $\psi=E_{\Delta}(H)\psi$\footnote{From now on, a
  $'$ will denote a derivative with respect to $y$.}.
Obviously $\Ga\ge 0$ is non-negative. The intuition is that if $\Ga$ goes to
$0$, then $\bigl(\psi,V_0\psi\bigr)$ is also small, whence $\psi$ is supported
in the bulk and cannot be an edge state, so that 
$\psi=E_{\Delta}(H)\psi$ is impossible. The problem is to estimate $\Ga$ in
terms of $\bigl(\psi,V_0\psi\bigr)$.

\begin{proof}[Proof of proposition \ref{thm:proposition1}]
Let $\eta=\dist(E,\sigma(H_0))$, so that for all
$\phi\in{\cal D}(H_0)$ in the domain of $H_0$,
$\norm{(E-H_0)\phi}\ge \eta\norm{\phi}$ holds.
The condition we put on $\delta$ is $\eta>\delta$. Then $E$ lies in the gaps
of the bulk Hamiltonian $H_0+V_d$.

Choose an $\epsilon>0$ with $\eta>\epsilon>\delta$ and a smooth ``cutoff''
function $j=j(y)$ with $1\ge j\ge 0$, $j(y)=1$
for $y\le b$ for some $b>0$ and $\sup(\abs{j(y)V(x,y)})\le\epsilon$. 
This is possible because of the assumptions on $V_0$ and because of 
$\abs{V_d}\le\delta <\epsilon$. (see figure \ref{fig:cutoff}).
\begin{figure}[ht]
\begin{center}
\psfrag{Vd}{$\displaystyle V_d\le\delta$}
\psfrag{eps}{$\epsilon$}
\psfrag{0}{$0$}
\psfrag{a}{$a$}
\psfrag{b}{$b$}
\psfrag{j}{$j$}
\psfrag{V0}{$V_0$} 
\psfrag{y}{$y$}
\includegraphics{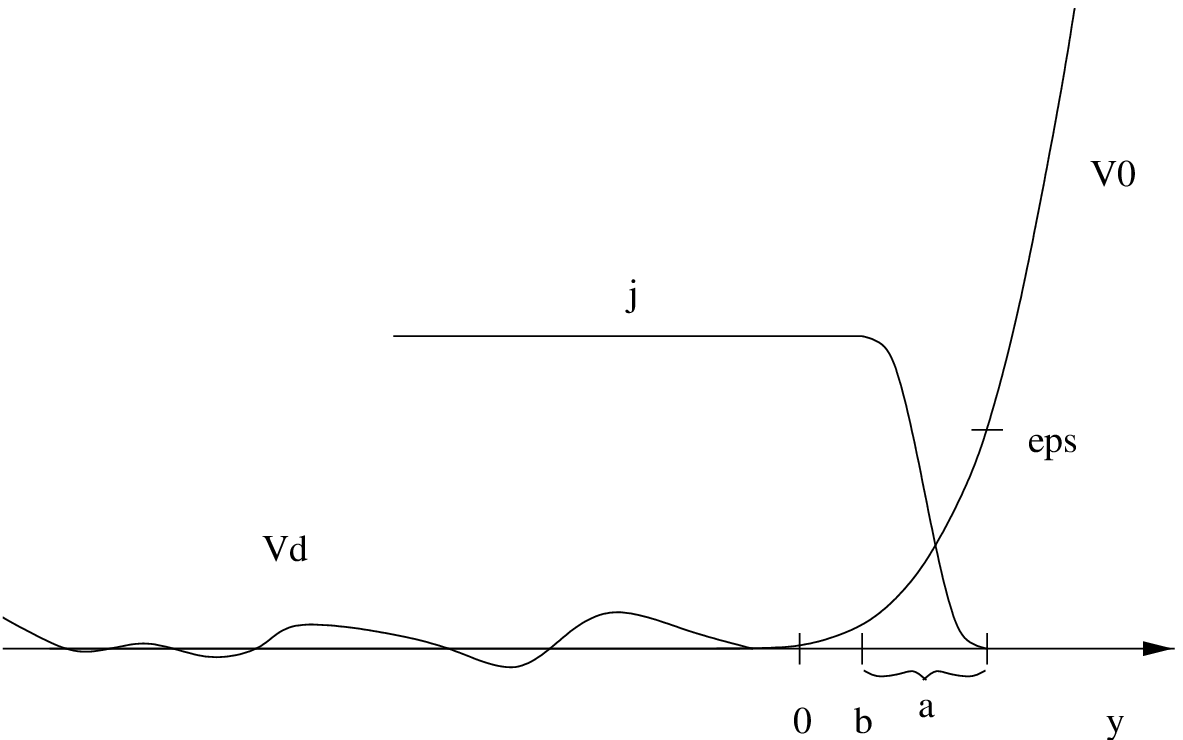}
\caption{The cutoff function $j$}
\label{fig:cutoff}
\end{center}
\end{figure}

We control $j$ and its derivatives by introducing the following finite
constants: 
\begin{equation}
\begin{split}
C_1&= \sup\{1/V_0'(y)\}:y\in \supp(1-j)\} \\
C_2&= \sup\{(j''(y))^2/V_0'(y):y\in \supp(1-j)\} \\
C_3&= \sup\{(j'(y))^2/V_0'(y):y\in\supp(1-j)\} \\
C_4&= \sup\{\abs{j'(y)}\}
\end{split}
\end{equation}
Keeping track of these different constants will later allow to determine the
dependence of the estimates on the steepness of the potential.
Now let $\Delta\ni E$ be an interval around $E$, and
$\psi=E_{\Delta}(H)\psi$. 
Then $\tilde\psi=j\psi\in {\cal D}(H_0)$ and the assumption on $E$ yields the
estimate
\begin{equation}
\eta\norm{j\psi} \le \norm{(E-H_0)j\psi}\le 
\norm{(E-H)j\psi}+ \underbrace{\norm{Vj\psi}}_{\displaystyle
\le\epsilon\norm{\psi}}. 
\label{eq:base}
\end{equation}
A bound of $\norm{j\psi}$ in the other direction is obtained from
\begin{equation}
\norm{(1-j)\psi}^2 = \bigl(\psi,(1-j)^2\psi\bigr) \le
C_1\Ga 
\end{equation}
and is
\begin{equation}
\norm{j\psi}\ge\norm{\psi}-\norm{(1-j)\psi} \ge \norm{\psi} -
C_1^{1/2}\Ga^{1/2}.  
\label{eq:hilf1}
\end{equation}
$Hj=jH-2i(p_y-A_y)j'+ j''$ yields
\begin{equation}
\norm{(E-H)j\psi} \le \norm{j(E-H)\psi}+ 2\norm{(p_y-A_y)j'\psi}
+\norm{j''\psi}. 
\label{eq:hilf2}
\end{equation}
The terms on the right hand side can be controlled in terms of $\abs{\Delta}$
and $\Ga$ as follows:
\begin{align}
\label{eq:hilf3}
\begin{split}
\norm{j''\psi}^2 &= \bigl(\psi,(j'')^2\psi\bigr) \le
C_2\Ga, \\
\norm{j(E-H)\psi} &\le \abs{\Delta}\norm{\psi},
\end{split} \\
\begin{split}
\norm{(p_y-A_y)j'\psi}^2 &= \bigl(\psi,j'(p_y-A_y)^2j'\psi\bigr) \\
 &\le \bigl(\psi,j'Hj'\psi\bigr) + \delta\bigl(\psi,(j')^2\psi\bigr)
\end{split}
\label{eq:hilf4}
\end{align}
because $V+\delta, (p_x-A_x)^2 \ge0$.
Using $j'Hj'=({j'}^2H+H{j'}^2)/2 + (j'')^2$, the first term on the right hand
side of \eqref{eq:hilf4} can be further estimated as
\begin{equation}
\label{eq:hilf5}
\begin{split}
\bigl(\psi,j'Hj'\psi\bigr)&= \frac{1}{2}\bigl(j'\psi,j'H\psi\bigr) + 
\frac{1}{2}\bigl(j'H\psi,j'\psi\bigr) + \bigl(\psi,(j'')^2\psi\bigr) \\
&\le \norm{j'\psi}\norm{j'H\psi} + \bigl(\psi, (j'')^2\psi\bigr).
\end{split} 
\end{equation}
Since 
\begin{align}
\norm{j'\psi}^2&\le C_3\Ga, \\
\norm{j'H\psi}&\le
\norm{j'(E-H)\psi}+ \norm{j'\psi} \le C_4\abs{\Delta}\norm{\psi}
+EC_3^{1/2}\Ga^{1/2},
\end{align}
equations \eqref{eq:hilf4} and \eqref{eq:hilf5} imply
\begin{equation}
\begin{split}
\norm{(p_y-A_y)j'\psi}^2 &\le \Ga[EC_3+ C_2+\delta C_3] \\
&\quad\qquad\qquad +\Ga^{1/2}C_3^{1/2}C_4\abs{\Delta}\norm{\psi} \\
\intertext{and thus}
\norm{(p_y-A_y)j'\psi}^{\phantom{2}} &\le \Ga^{1/2}[EC_3+ C_2+\delta
C_3]^{1/2} \\ 
&\quad\qquad\qquad +\Ga^{1/4}C_3^{1/4}C_4^{1/2}\abs{\Delta}^{1/2}
\norm{\psi}^{1/2}
\end{split} 
\end{equation}
We now combine \eqref{eq:base}, \eqref{eq:hilf1}, \eqref{eq:hilf2},
\eqref{eq:hilf3} and the last inequality to:
\begin{equation}
\begin{split}
\eta\Bigl(\norm{\psi}-C_1^{1/2}\Ga^{1/2}\Bigr)& \le \eta\norm{j\psi} \\ 
&\le \norm{(E-H)j\psi} + \epsilon\norm{\psi} \\
&\le \abs{\Delta}\norm{\psi} + \epsilon\norm{\psi} + C_2^{1/2}\Ga^{1/2} +\\
&\quad\qquad + 2\Ga^{1/2}\bigl[(E+\delta)C_3+ C_2\bigr]^{1/2} \\
&\quad\qquad\qquad + 2\Ga^{1/4}C_3^{1/4}C_4^{1/2}\abs{\Delta}^{1/2} 
\norm{\psi}^{1/2} 
\end{split}
\end{equation}
Abbreviating $D_1= C_2^{1/2} + 2\bigl[(E+\delta)C_3+
C_2\bigr]^{1/2}\!\!,\,D_2=2C_3^{1/4}C_4^{1/2}$ and $ D_3=C_1^{1/2}$,
the conclusion is that for all $\psi=E_{\Delta}(H)\psi$:
\begin{equation}
\begin{split}
(\eta-\abs{\Delta}-\epsilon)\norm{\psi} &
\le \Ga^{1/2}(D_1+ \eta D_3)+ D_2\Ga^{1/4} \abs{\Delta}^{1/2}\norm{\psi}^{1/2}
\\
&\le 2\Ga^{1/2}(D_1+\eta D_3) + (\lambda-1) \abs{\Delta} \norm{\psi},   
\end{split}
\end{equation}
where $\lambda-1= D_2^2/4(D_1+\eta D_3)$.
Since $\eta-\epsilon>0$, and by making $\abs{\Delta}$ small enough, one
finally gets:  
\begin{equation}
\Ga\ge \left[\frac{\eta-\lambda\abs{\Delta}-\epsilon}{2(D_1+\eta
    D_3)}\right]^2 \norm{\psi}^2 =: \tilde\alpha\norm{\psi}^2 
\label{eq:alphatilde}
\end{equation} 
\end{proof}

\begin{proof}[Proof of theorem \ref{thm:mourreest1}]
The missing piece is the estimate of $\partial_y\!V_d=-[V_d,ip_y]$ on the
energy interval $\Delta$. 
\begin{equation}
\begin{split}
\abs{\bigl(\psi,[V_d,ip_y]\psi\bigr)}&=\abs{\bigl(\psi,[V_d,i(p_y-A_y)] 
\psi\bigr)} \\ 
&=\abs{\bigl(V_d\psi,(p_y-A_y)\psi\bigr) -
\bigl((p_y-A_y)\psi,V_d\psi\bigr)} \\
&\le 2\delta\norm{\psi}\norm{(p_y-A_y)\psi}
\end{split}
\end{equation}
For $\psi=E_{\Delta}(H)\psi$ with $\Delta\ni E$,
\begin{align}
\begin{split}
\norm{(p_y-A_y)\psi}^2 &= \bigl(\psi,(p_y-A_y)^2\psi\bigr) \le
\bigl(\psi,(H+\delta)\psi\bigr)\\ 
&\le (E+\abs{\Delta}+\delta)\norm{\psi}^2
\end{split}
\intertext{so that}
\abs{\bigl(\psi,[V_d,ip_y]\psi\bigr)} &\le 2\delta(E+\abs{\Delta}+\delta)^{1/2}
\norm{\psi}^2. 
\end{align}
With the additional condition on $\delta$, 
\begin{equation}
2\delta(E+\abs{\Delta}+\delta)^{1/2}< \tilde\alpha,
\label{eq:zusatzbed}
\end{equation}
the proof is complete.
\end{proof}



We now discuss the dependence of the estimate on the assumptions about the
disorder and edge potentials.

As mentioned above, one can relax the constraints on $\delta$ by introducing
another constant $\delta'$ and imposing $\abs{V'_d}\leq\delta'<\tilde\alpha$
with $\tilde\alpha$ as in \eqref{eq:alphatilde}. It is actually enough if
$V'_d$ is small near the edge, as one can easily show in a manner similar to
the above by introducing a partition of unity separating the regions where
$V_d$ is smooth from those where it is rougher. The length scale is 
of course set by
the cyclotron length $l_c=1/\sqrt{B}$: If $V_d$ varies strongly on a scale of
$l_c$, it is better to use $\delta\sqrt{E}$ as in \eqref{eq:zusatzbed} to
control $\abs{V'_d}$. If $V_d$ is smooth on this scale, the use of a $\delta'$
is more appropriate. 

The Mourre estimate can thus be established for the following classes of
disorder potentials:
\begin{itemize}
\setlength{\parskip}{0cm}
\item If $V_d$ is small enough, no smoothness assumptions are needed. 
\item If the first derivative is small near the edge, $V_d$ itself can be a
  little larger.
\end{itemize}

We now turn to the dependence of the estimates on the edge potential.
$\tilde\alpha$ as defined in \eqref{eq:alphatilde} depends 
not only on the disorder potential via $\delta$, but also on $V_0$ via the
constants in the denominator, which are constrained by
$\abs{jV}<\epsilon$: Assume
$V_0$ increases from $0$ to $\epsilon$ on a length scale $a$. Then
$j$ must go from $1$ to $0$ on this scale (see figure \ref{fig:cutoff}),
and for small $a$, the constants vary as:
\begin{align}
C_1 &\sim \frac{a}{\epsilon} &
C_2 &\sim \frac{1}{\epsilon a^3}& 
C_3 &\sim \frac{1}{\epsilon a} & 
D_1^2 &\sim \frac{1}{\epsilon a^3} & 
D_3^2 &\sim \frac{a}{\epsilon}
\label{eq:skalen}
\end{align}

Together with \eqref{eq:alphatilde}, small $a$ or a steep potential implies
$\tilde\alpha\sim(\eta-\epsilon)^2\epsilon a^3$. 
A steeper edge thus seems to allow less disorder.
This problem is unexpected, since in the classical case also a hard wall leads
to extended edge states. Using dimensional analysis, one can argue that any
direct estimate of $\Ga$ in terms of $\eta$, which is roughly the same as
$\bigl(\psi,V_0\psi\bigr)$,  will have a dependence on $a$ that makes it
fail when $a$ tends to zero. 

Before we return to this problem in the next section by analyzing the problem
with Dirichlet boundary conditions, we indicate what must be changed in the
case of a bounded edge potential.

\begin{figure}[ht]
\begin{center}
\psfrag{j1}{$j_1$}\psfrag{j2}{$j_2$}\psfrag{j3}{$j_3$}
\psfrag{E0}{$E_0$}
\psfrag{V0}{$V_0$}
\psfrag{eps}{$\epsilon$}
\psfrag{y}{$y$}
\psfrag{0}{$0$}
\includegraphics{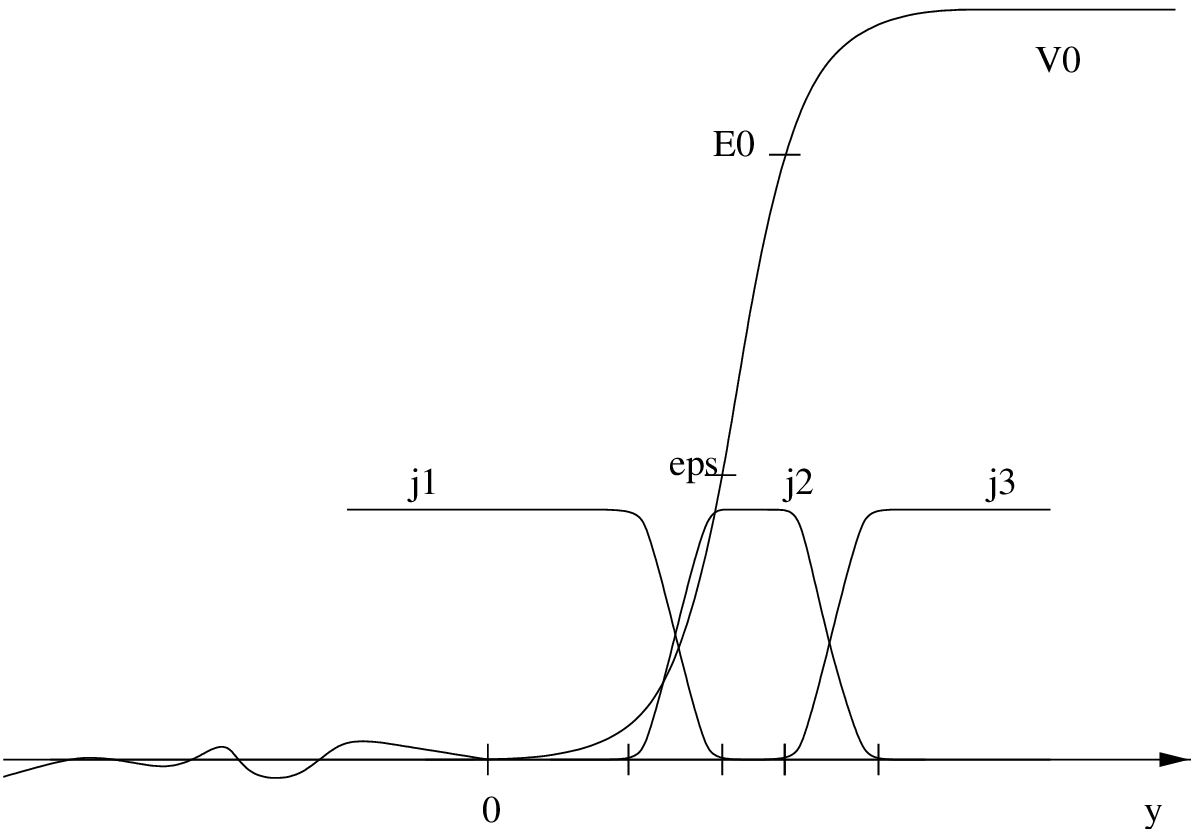}
\caption{Partition of unity for bounded $V_0$}
\label{fig:potentialflachweitdraussen}
\end{center}
\end{figure}

We assume that the edge potential levels off above some height $E_0$, but
still with $V_0'\ge 0$.
Let $\eta=\min(\dist(E,\sigma(H_0)),E_0-E)$, and choose
$\delta<\epsilon<\eta$ as above. Introduce a partition of unity according to 
figure \ref{fig:potentialflachweitdraussen}, satisfying
 $\sup(\abs{j_1V})\le\epsilon$ and $V_0\ge E_0> E$ on $\supp j_3$. The
 condition $\inf\{V_0'(y):y\ge b\}>0$ for all $b>0$ is replaced by the
 condition $\inf\{V_0'(y):y\in\supp(j_2)\}>0$.

From the proof of proposition \ref{thm:proposition1}, we know that for small
$\abs{\Delta}$,
\begin{align}
\begin{split}
\eta\norm{j_1\psi} &\le \norm{(E-H_0)j_1\psi} \\
&\le \lambda\abs{\Delta}\norm{\psi} + \epsilon\norm{j_1\psi} + C\Ga^{1/2}.
\label{eq:boundedm}
\end{split}
\intertext{The constants $C$ and $\lambda$ might change subsequently, but are
  independent of $\psi$, $\delta$, and $\epsilon$. Since
  $\eta-\epsilon\le V_0-E+V_d$ on $\supp(j_3)$,}
\begin{split}
(\eta-\epsilon)\norm{j_3\psi}^2 &= (\eta-\epsilon)\bigl(\psi,j_3^2\psi\bigr)\\
&\le \bigl(\psi,j_3(H-E)j_3\psi\bigr) \\
&\le \norm{j_3\psi}\norm{j_3(H-E)\psi} + \bigl(\psi,(j_3')^2\psi\bigr) \\
&\le \norm{j_3\psi} \abs{\Delta}\norm{\psi} + \norm{j_3\psi} C
\Ga^{1/2}. 
\end{split} 
\intertext{Therefore,}
\eta\norm{j_3\psi}&\le \abs{\Delta}\norm{\psi} + \epsilon\norm{j_3\psi} + 
C\Ga^{1/2}. 
\label{eq:boundedn}
\intertext{Together with}
\norm{\psi} &\le \norm{j_1\psi} + \norm{j_2\psi} +\norm{j_3\psi}, \\
\norm{j_2\psi} &\le C \bigl(\psi,V'_0\psi\bigr)^{1/2},
\end{align}
equations \eqref{eq:boundedm} and \eqref{eq:boundedn} yield
\begin{equation}
\eta\norm{\psi} \le \epsilon\norm{\psi} + \lambda\abs{\Delta}\norm{\psi} +
C\Ga^{1/2} 
\end{equation}
so that one gets an estimate of $\Ga$ as above.

\begin{figure}[h]
\begin{center}
\psfrag{ver}{impurity}
{\includegraphics{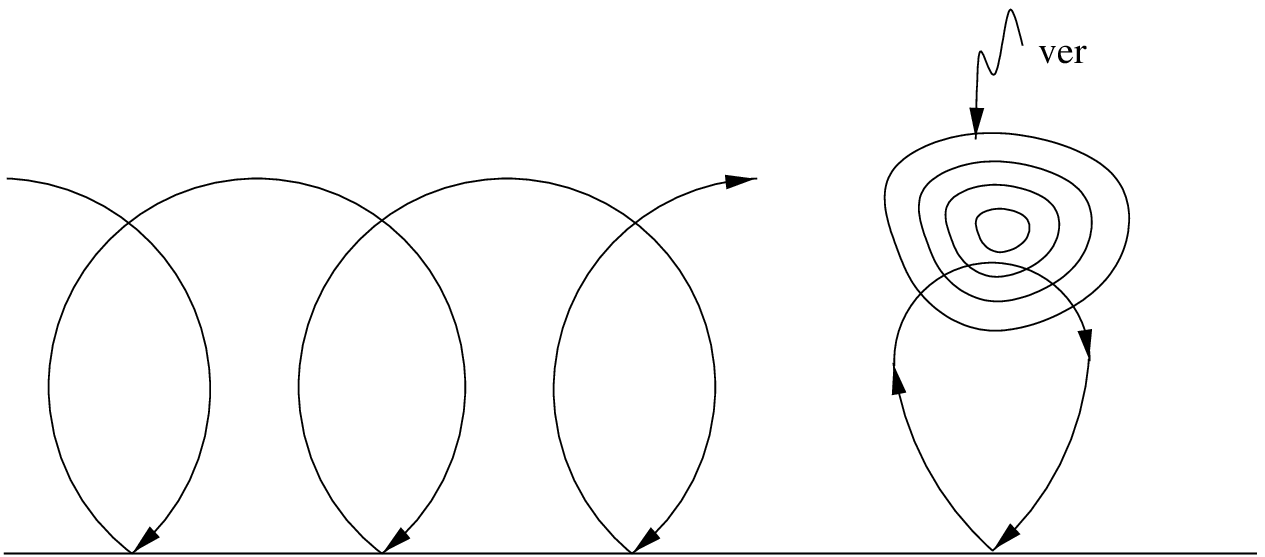}}
\caption{Classical ``marginally edge'' state and its localization by an
  impurity} 
\label{fig:locedge}
\end{center}
\end{figure}

Since it is not possible to establish the Mourre estimate for the whole
spectral gaps of the bulk Hamiltonian,
it is natural to ask whether it is possible that not all
edge states are extended. This question is difficult as a quantum
mechanical problem, but some insight can be obtained by looking
at the corresponding classical picture. Here edge and bulk states can be
distinguished, not merely by their energy, but according to the criterion
whether or not the orbit hits the edge. It is then very easy to imagine a
localized edge state, see fig. (\ref{fig:locedge}). 
 
\subsection{Dirichlet boundary conditions} 
\label{sec:diric}
The somewhat more technical analysis of the half-plane with Dirichlet
boundary conditions in this section is included to support the expectation
that spectrum in between Landau levels is still absolutely continuous
in the limit of an infinitely steep wall, although our estimates
in section \ref{sec:edge} are not uniform in the steepeness of the wall.
We again restrict ourselves to one edge, that is, we consider $H$ on the
half-space $y<0$ and require the wave functions to vanish at $y=0$.
This section is organized as follows: We first define the necessary spaces and
operators, and compute the commutator $[H,i\Pi]$. Postponing the resolution of
the appearing difficulties, we then prove the Mourre estimate. At the end of
the section, we explain why this estimate indeed implies absolute continuity
of spectrum. 

\subsubsection{Domains and operators}

The ``physical'' Hilbert space is $\Hi ^- =L_2\bigl(\R\times
(-\infty,0)\bigr)=L_2(\R\times\R_-)=L_2(\R^2_-)$. The corresponding
scalar product is denoted by $(\,\cdot\,,\,\cdot\,)_-$, the norm by
$\norm{\,\cdot\,}_-$. We will also need the Hilbert-space $\Hi=L_2(\R^2)$,
whose scalar product will be denoted simply by $(\,\cdot\,,\,\cdot\,)$.

As is well known, the momentum operator $p_y=-i\partial_y$ cannot be defined
as a self-adjoint operator on $\Hi^-$. Although $p_y$ is, for example,
symmetric with domain $\De(p_y)=C^{\infty}_c(\R\times\R_-)$, the deficiency
indices are $(\infty,0)$, so that there cannot be a self-adjoint extension.
The operator $\Pi=p_y+Bx$, which will be used as the conjugate operator
is defined as the closure of the operator $-i\partial_y+Bx$ on
$C^{\infty}_c(\R\times\R_-)$. 

In contrast, the other component of the kinematical momentum, $p_x+By$, 
is a well-defined self-adjoint operator. 
The Hamiltonian $H$ can be defined as the Friedrichs extension of a
quadratic form. For $\phi,\psi\in C^{\infty}_c(\R\times\R_-)$, let 
\begin{equation}
h(\phi,\psi)= \bigl(p_y\phi,p_y\psi)_- + 
\bigl((p_x+By)\phi,(p_x+By)\psi\bigr)_- +\bigl(\phi,V_d\psi\bigr)_-. 
\end{equation}
This form is bounded from below, and yields a unique self-adjoint $H$ with
$h(\phi,\psi)=\bigl(\phi,H\psi\bigr)_-$ for all
$\phi,\psi\in C^{\infty}_c(\R\times\R_-)$.    

Another equivalent characterization of $H$ is:
\begin{equation}
H= p_y^{\dagger}p_y + (p_x+By)^2 + V_d,
\end{equation}
where $p_y$ is the closure of $-i\partial_y$ on $C_c^{\infty}(\R^2_-)$.

To control the regularity of $H$ with respect to $\Pi$, we introduce the
following scale
of Hilbert spaces associated with $H$: For $s\in\R$, let $\Hi^-_s$ be the
domain of 
$\abs{H}^{s/2}$ equipped with the graph-norm. Note that because of the presence of
the magnetic field, those $\Hi^-_s$ are not the same as the usual
Sobolev spaces for the half-space,
$\So{s}(\R^2_-)$, which are defined by means of the Laplacian. But in Landau
gauge the magnetic field enters only in the $y$-direction, so that
``locally'', $\Hi_s^-$ and $\So{s}(\R^2_-)$ coincide. 

In this section, $H_0 = p_y^2+(p_x+By)^2$ on $\Hi=L_2(\R^2)$, i.e. without
boundary conditions.

Dirichlet boundary conditions without disorder is again simple. After
separating the $y$-coordinate of the guiding center, the remaining
one-dimensional problems with boundary condition can be solved in terms of
particular hypergeometric functions. One thereby obtains the functions
$E_n(k)$. The spectrum of $H$ is the interval
$[B,\infty)$ and absolutely continuous. A reasonable random disorder potential
changes the location of the spectrum only marginally (see appendix
\ref{sec:app2}).

\subsubsection{The commutator}

To compute the commutator
\begin{equation}
[H,i\Pi]= [p_y^{\dagger}p_y+(p_x+By)^2+V_d,i(p_y+Bx)],
\end{equation}
first note that $[(p_x+By)^2,p_y+Bx]=0$, as does
$[p_y^{\dagger}p_y,Bx]=0$. It is also simple to see
$[V_d,i(p_y+Bx)] = -\partial_y\!V_d$.
For $[p_y^{\dagger}p_y,p_y]$, let 
$\phi,\;\psi\in \De(H)$. Then as a quadratic form
\begin{equation}
\bigl(\phi,[p_y^{\dagger}p_y,ip_y]\psi\bigr)_- =
\bigl(p_y^{\dagger}p_y\phi,ip_y\psi\bigr)_- -
\bigl(p_y^{\dagger}\phi,ip_y^{\dagger}p_y\psi\bigr)_-.  
\label{eq:quadraform}
\end{equation}
$p_y\phi$ is not anymore in the domain of $p_y$, but it is
in the domain of $p_y^{\dagger}$ and with the representation 
$p_y^{\dagger}=-i\partial_y$ and $p_y^{\dagger}\phi=p_y\phi$, one has
\begin{equation}
\phantom{=} -\bigl(\partial_yp_y\phi,p_y\psi\bigr)_- - 
\bigl(p_y\phi,\partial_yp_y\psi\bigr)_-.
\end{equation}
After a partial integration, only the boundary term
\begin{equation}
\phantom{=} -\int\de{x}\, \bigl(p_y\phi(x,0)\bigr)^* p_y\psi(x,0)  
\label{eq:randterm}
\end{equation}
is left. The commutator thus is
\begin{equation}
\begin{split}
\bigl(\phi,[H,i\Pi]\psi\bigr)_- &=-\bigl(p_y\phi,p_y\psi\bigr)_0
-\bigl(\phi,\partial_y\!V_d\psi\bigr)_- \\
&= -\bigl(\phi',\psi'\bigr)_0 -\bigl(\phi,\partial_y\!V_d\psi\bigr)_-,
\end{split}
\end{equation}
where $-\bigl(\phi',\psi'\bigr)_0$ stands for \eqref{eq:randterm}.

From the theory of Sobolev spaces and boundary value problems in 
partial differential equations (see \cite{LM}), it is known that if
$f\in\So{s}(\R\times\R_-)$ and  
$s>3/2$, then the restriction of the first $y$-derivative of $f$ on the
boundary of the half-space is in
$\So{s-3/2}(\R)$ and that the map
$f\mapsto\partial_yf\restrict _{y=0}$ is continuous in the corresponding
norms. (\cite{LM}, chapter 1, section 7.2). 
Since ``locally'' $\Hi^-_s$ coincides with $\So{s}(\R\times\R_-)$, this shows
that for every $\epsilon>0$, the first commutator is a bounded operator from 
$\Hi^-_{3/2+\epsilon}$ to $\Hi^-_{-3/2-\epsilon}$. That the magnetic field
does not invalidate this continuity can be proved by adapting the 
arguments given in \cite{LM}. We note that this implies that for every
$0\le\epsilon < 1/2$, the expression
\begin{equation}
(H+1)^{1-\epsilon/2}[H,i\Pi](H+1)^{1-\epsilon/2}
\label{eq:bounded}
\end{equation}
is a well-defined and bounded operator on $\Hi^-$.

The most obvious problem with the application of Mourre theory in the
present case is that $\Pi$ is not self-adjoint. 
The other question is whether the commutator is sufficiently regular.
We will return to those issues after we have established the Mourre
estimate, which is the content of the next theorem. 

\subsubsection{Mourre estimate}

\begin{theorem}
\label{thm:mourreest2}
Assume $E\notin\{(2n+1)B,n\in\N\}$. Then there exist constants $\delta$,
$\delta'$, and $\delta''$ such that if the 
disorder potential satisfies $|V_d| \leq\delta$, 
$\abs{\partial V_d}\leq\delta'$, and $\abs{\partial^2 V_d}\le \delta''$ 
there is an open interval $\Delta\ni E$ and a
positive constant $\alpha$ with
\begin{equation}
-E_{\Delta}(H)\left[ H,i\Pi \right] E_{\Delta}(H) \geq \alpha E_{\Delta}(H)  
\end{equation}
\end{theorem}
 
As in section \ref{sec:edge}, the Mourre estimate (Theorem
\ref{thm:mourreest2}) will follow once we show that the part of the
commutator arising from the edge is positive. The part of the commutator
arising from the disorder potential is controlled by $\delta'$. 

The assumption on the derivatives of the disorder potential could be weakened
considerably within the present approach. Also, De Bi\`evre and Pul\'e
\cite{DBP} have reported a Mourre estimate for the hard wall that avoids such 
an assumption. 

\begin{prop}
Under conditions as in Theorem \ref{thm:mourreest2}, there is an
$\tilde\alpha>0$ and a $\Delta\ni E$, such that for all $\psi$ with
$\psi=E_{\Delta}(H)\psi$: 
\begin{equation}
\Gamma_{\psi} := \int \abs{\partial_y\psi(x,0)}^2 \de{x} \ge
\tilde\alpha\norm{\psi}^2_-
\end{equation}
\label{thm:proposition}
\end{prop}

One proof of this proposition uses a cutoff function,
similarly to section \ref{sec:edge}, and can be found in appendix
\ref{sec:app3}. In a different proof, presented here, $\psi$ is continued into 
the half-space $y>0$.

\begin{proof}[Proof of proposition \ref{thm:proposition}]
Let $\psi=E_{\Delta}(H)\psi$. Decompose $\psi'(x,0)\in L_2(\R)$ in 
Fourier components:
\begin{equation}
\psi'(x,0)= \int \frac{\de{k}}{2\pi} \hat\psi'(k,0) e^{ikx}.
\end{equation}
For every Fourier component, introduce an extension $j_k(y)$ that is
sufficiently smooth and satisfies the conditions: $j_k(0)=0$, $j'_k(0)=1$.
Define the extension of $\psi$ by
\begin{equation}
\tilde\psi(x,y)= 
\begin{Cases}
\text{\raisebox{0.5em}{$\psi(x,y)$}} & \text{\raisebox{0.5em}{for $y<0$}} \\
\text{\raisebox{-0.2em}{$\displaystyle
\int\frac{\de{k}}{2\pi} j_k(y) \hat\psi'(k,0) e^{ikx}$}} &
\text{\raisebox{-0.2em}{for $y\ge0$}} 
\end{Cases}
\end{equation}
This extension is obviously $C^1$ and 
$\tilde\psi\in{\cal D}(H_0)$. Letting  
$\eta=\dist(E,\sigma(H_0))$ and writing
$\nnorm{\,\cdot\,}_-=\nnorm{\chi_{\R\times\R_-}\,\cdot\,}$ and
$\nnorm{\,\cdot\,}_+=\nnorm{\chi_{\R\times\R_+}\,\cdot\,}$, one has:
\begin{equation}
\begin{split}
\eta\norm{\psi}_- = \eta\norm{\psi} &\le\eta\nnorm{\tilde\psi} \le
\nnorm{(E-H_0)\tilde\psi} \\ &
\le \norm{(E-H)\psi}_- + \nnorm{(E-H_0)\tilde\psi}_+ +\delta\norm{\psi}_-
\label{eq:grundfort}
\end{split}
\end{equation}
Now
\begin{multline}
\norm{\smash{(E-H_0)\tilde\psi}}_+^2 =\\
\begin{split}
&= \int_{y\ge0}\de{x}\de{y} 
\left\lvert \int\frac{\de{k}}{2\pi} \Bigl(E-\bigl(p_y^2+(p_x+By)^2\bigr)\Bigr)
\Bigl(j_k(y) \hat\psi'(k,0) e^{ikx}\Bigr)
\right\rvert^2  \\
&= \int_{y\ge0}\de{x}\de{y} \left\lvert\int\frac{\de{k}}{2\pi}
  e^{ikx}\hat\psi'(k,0) 
  \Bigl(E-\bigl(p_y^2 + (k+By)^2\bigr)\Bigr)j_k(y)\right\rvert^2 \\ 
&= \int\frac{\de{k}}{2\pi} \biggl(\abs{\hat\psi'(k,0)}^2 
  \int_{y\ge0}\de{y}\left\lvert\Bigl(E-\bigl(p_y^2 +
  (k+By)^2\bigr)\Bigr)j_k(y)\right\rvert^2\biggr).
\end{split}
\end{multline}
The last $y$-integral has dimensions of an inverse length, and by choosing,
for example, 
$j_k(y)= ye^{-y^2l(k)^2}$ with $l(k)=\abs{k}+\sqrt{B}$, it is bounded
by a constant times $\abs{k}+\sqrt{B}$.
Then equation \eqref{eq:grundfort} implies:
\begin{equation}
\eta\norm{\psi}_- \le \bigl(\abs{\Delta}+\delta\bigr)\norm{\psi}_- 
+ \biggl(C \int \frac{\de{k}}{2\pi}
\abs{\hat\psi'(k,0)}^2 \Bigl(\abs{k}+ \sqrt{B}\Bigr)
\biggr)^{1/2}\negthickspace\negthickspace,
\end{equation}
and because of
\begin{equation}
\int \frac{\de{k}}{2\pi} \nabs{\hat\psi'(k,0)}^2 = 
\int \de{x} \abs{\psi'(x,0)}^2 = \Gamma_{\psi},
\end{equation}
one is left with estimating the integral
\begin{equation}
\int \frac{\de{k}}{2\pi} \nabs{\hat\psi'(k,0)}^2 \abs{k}
\end{equation}
in terms of $\Gamma_{\psi}$. Now,
\begin{equation}
\begin{split}
\biggl(\int \frac{\de{k}}{2\pi} \nabs{\hat\psi'(k,0)}^2 \abs{k}\biggr)^2 &\le
\biggl(\int \frac{\de{k}}{2\pi} \nabs{\hat\psi'(k,0)}^2\biggr)
\biggl(\int \frac{\de{k}}{2\pi} \nabs{\hat\psi'(k,0)}^2 \abs{k}^2\biggr) \\
& = \Gamma_{\psi} \;\bigl(p_y\psi, p_x^2\,p_y\psi\bigr)_0. 
\end{split}
\end{equation}
Then, because of $\psi(x,0)=0$,
\begin{equation}
\begin{split}
\bigl(p_y\psi, p_x^2\,p_y\psi\bigr)_0 
&= \bigl(p_y\psi,(p_x+By)^2\,p_y\psi\bigr)_0  \\
&= \bigl(p_y\psi, p_y\,(p_x+By)^2\psi\bigr)_0.  \\
\intertext{Using the commutator $[p_y^{\dagger}p_y,ip_y]$ (see equations
  \eqref{eq:quadraform} to \eqref{eq:randterm}) and
  $p_y^{\dagger}\psi=p_y\psi$, this can be written as} 
\bigl(p_y\psi, p_x^2\,p_y\psi\bigr)_0 
&= -\bigl(p_y^{\dagger}p_y\psi,ip_y(p_x+By)^2\psi\bigr)_-+ 
   \bigl(p_y\psi,ip_y^{\dagger}p_y(p_x+By)^2\psi\bigr)_-, \\
   \displaybreak[0]  
\intertext{and estimated by}
\abs{\bigl(p_y\psi, p_x^2\,p_y\psi\bigr)_0} 
&\le\abs{\bigl(p_y^{\dagger}p_y\psi, p_y\,(p_x+By)^2\psi\bigr)_-} + 
\abs{\bigl(p_y\psi,p_y^{\dagger}p_y\, (p_x+By)^2\psi\bigr)_-} \\
&\le \abs{\bigl(p_y^{\dagger}p_y^{\dagger}p_y\psi,(p_x+By)^2\psi\bigr)_-} +
\abs{\bigl((p_x+By)^2\psi,p_y^{\dagger}p_y^{\dagger}p_y\psi\bigr)_-} \\
&\phantom{=} +
\abs{\bigl(\psi,\bigl[p_y^{\dagger}p_y^{\dagger}p_y,(p_x+By)^2\bigr]
  \psi\bigr)_-}\\[0.2cm]   
&= \abs{\bigl(p_y^{\dagger}p_y^{\dagger}p_y\psi,(p_x+By)^2\psi\bigr)_-} +
\abs{\bigl((p_x+By)^2\psi,p_y^{\dagger}p_y^{\dagger}p_y\psi\bigr)_-} \\ 
&\phantom{=} +2B\bigl\lvert\bigl(\psi,
  \bigl(p_y^{\dagger}p_y\,(p_x+By)+ p_y^{\dagger}(p_x+By)p_y \\
&\phantom{=} \qquad\qquad\qquad\qquad\qquad\qquad\qquad 
+ (p_x+By)\,p_y^{\dagger}p_y\bigr)\psi\bigr)_-\bigr\rvert 
\\[0.2cm] 
&\le 2\norm{p_y^{\dagger}p_y^{\dagger}p_y\psi}_-\norm{(p_x+By)^2\psi}_- \\
&\phantom{\le} +
 6B\norm{p_y^{\dagger}p_y\psi}_-\norm{(p_x+By)\psi}_- 
+ 2B^2\norm{p_y\psi}_-\norm{\psi}_-. \\[0.2cm]
\end{split}
\end{equation}
Estimates for powers of $p_y$ and $p_x+By$ by the Hamiltonian are proved in
lemma \ref{thm:kato}, and yield 
\begin{equation}
\bigl(p_y\psi, p_x^2\,p_y\psi\bigr)_0 \le C \norm{\psi}_-^2,
\end{equation}
where $C$ depends on $B$, $E$, $\abs{\Delta}$, and the bounds on the disorder
potential and its derivatives.

\begin{equation}
\biggl(\int \frac{\de{k}}{2\pi} \abs{\hat\psi'(k,0)}^2 \abs{k}\biggr)^2
\le \Gamma_{\psi} C\norm{\psi}^2_-
\end{equation}
now yields the desired estimate
\begin{equation}
\eta\norm{\psi}_- \le \bigl(\abs{\Delta}+\delta\bigr)\norm{\psi}_- + C
\Bigl(\Gamma_{\psi} +   
\sqrt{\Gamma_{\psi}}\norm{\psi}_- \Bigr)^{1/2}
\negthickspace\negthickspace,
\end{equation}
from which the claim follows immediately.
\end{proof}

\begin{lemma}
\label{thm:kato}
There is a constant $C$ that depends on $E$, $\abs{\Delta}$, and on $\delta$,
$\delta'$, and $\delta''$ such that if $\psi=E_{\Delta}(H)\psi$, estimates
of the form $\norm{p_y\psi}_-\le C\norm{\psi}_-$, $\norm{(p_x+By)\psi}_-\le
C \norm{\psi}_-$, $\norm{p_y^{\dagger}p_y\psi}_- \leq C\norm{\psi}_-
$, and $\norm{p_y^{\dagger}p_y^{\dagger}p_y\psi}_- \leq C
\norm{\psi}_-$ hold. 
\end{lemma}

\begin{proof}
The first two estimates are trivial. For $\norm{p_y^{\dagger}p_y\psi}_-$,
observe that the expressions
\begin{equation}
\begin{split}
p_y^{\dagger}p_y(p_x+By)^2 + (p_x+By)^2p_y^{\dagger}p_y + 2B^2 = 
2p_y^{\dagger} (p_x+By)^2 p_y \ge 0 \\
p_y^{\dagger}p_y(V_d+\delta) + (V_d+\delta)p_y^{\dagger}p_y + \partial_y^2V_d
=  2p_y^{\dagger} (V_d+\delta) p_y \ge 0 \\
(p_x+By)^2(V_d+\delta) + (V_d+\delta)(p_x+By)^2 + \partial_x^2V_d = \\
2(p_x+By)(V_d+\delta)(p_x+By)\ge 0 
\end{split}
\end{equation}
are positive. Therefore
\begin{equation}
\begin{split}
\norm{p_y^{\dagger}p_y\psi}_-^2 &=
\bigl(p_y^{\dagger}p_y\psi,p_y^{\dagger}p_y\psi\bigr)_- \\
&\le \bigl((H+\delta)\psi,(H+\delta)\psi\bigr)_- + 2B^2\norm{\psi}_-^2 +
2\delta''\norm{\psi}_-^2 \\ 
&\le \bigl((E+\abs{\Delta}+\delta)^2+2B^2+ 2\delta''\bigr)\norm{\psi}_-^2
\end{split}
\end{equation}
This shows $\norm{p_y^{\dagger}p_y\psi}_- \leq C\norm{\psi}_-$. The 
estimate for $\norm{p_y^{\dagger}p_y^{\dagger}p_y\psi}_-$ can be obtained
similarly, but the calculations are somewhat longer, and we omit them here.
\end{proof}

\subsubsection{Absolutely continuous edge spectrum}

In the last part of this section, we explain how the extensions of Mourre
theory allow us to conclude absolute continuity of spectrum from the
Mourre estimate. 

In references \cite{sahbani} and \cite{ABG}, the regularity properties of a
self-adjoint operator $H$ on a Hilbert-space $\Hi$
with respect to another self-adjoint operator $A$ are defined via the unitary
group 
generated by $A$. A bounded operator $H\in\Be(\Hi)$ is said to be of
regularity class 
$C^1(A)$ if the map
\begin{equation}
\tau\mapsto\e^{iA\tau}H\e^{-iA\tau} =: \We(\tau)H
\end{equation}
is continuously differentiable at $\tau=0$, in the strong operator norm 
on $\Be(\Hi)$. This is equivalent to the existence of the commutator $[H,iA]$
as a bounded operator in $\Hi$.
One can also introduce the regularity classes
$\Ce^{1+s}(A)$ for $0<s<1$:
$H\in\Ce^s(A)$ if  $\We(\tau)H$ is differentiable
and if the derivative is $s$-H\"older continuous at $\tau=0$. 
A class $\Ce^{1+0}$ is
defined via Dini continuity of the derivative.
If $H$ is unbounded, the regularity classes are defined via the resolvent
$(H-z)^{-1}$.
For the proof of absolute continuity of spectrum on the basis of the
positivity of a commutator it suffices to show that
$H\in \Ce^{1+0}(A)$ (see \cite{sahbani}).

To reduce the half-plane problem with Dirichlet boundary conditions to a
situation to which the above can be applied, we proceed in steps:
\begin{enumerate}
\setlength{\leftmargini}{0cm}
\setlength{\labelwidth}{0cm}
\setlength{\parskip}{0cm}
\item For convenience, we now also set the magnetic field $B=1$.
For small $V_d$ (compared to the magnetic field),
$-1\notin\sigma(H)$. Then the resolvent $R=(H+1)^{-1}$ is a bounded operator
on $\Hi^-$ and can be extended by $0$ to $\Hi$. The extension will
be denoted by $\tR$. For every $\psi\in\Hi$, $\tR\psi$ is in the domain of $H$
and therefore vanishes on the boundary $y=0$. 
The non self-adjoint operator $\Pi=p_y+x$ on $\Hi^-$ has a self-adjoint
counterpart $\tPi=\tilde p_y+x$ on $\Hi$, where $\tilde p_y$ is defined in the
usual manner as the closure of $-i\partial_y$ on $C^{\infty}_c(\R^2)$. If we
denote by $P_-$ the projection from $\Hi$ onto $\Hi^-$,
we have the relationship $P_-\tPi=\Pi P_-$ and $P_- H_0 = H P_-$ between
operator on $\Hi^-$ and $\Hi$. Thus $\tPi$ is not an extension of $\Pi$, which
would read $P_- \Pi P_- = P_-\tPi P_-$. We also note that $(H+1)\tR= \tR
(H_0+1)=P_-$. 


\item For $\phi,\psi \in \De(\tPi)$, we compute the commutator $[\tR,\tPi]$ as
  a quadratic form:
\begin{equation}
\begin{split}
\bigl(\phi,[\tR,i\tPi]\psi\bigr) &= i\bigl(\tR\phi,\tPi\psi\bigr)
-i\bigl(\tPi\phi,\tR\psi\bigr) \\
&= i\bigl(\tR\phi,P_-\tPi\psi\bigr) - i\bigl(P_-\tPi\phi,\tR\psi\bigr) \\
&= i\bigl(RP_-\phi,\Pi P_-\psi\bigr) - i\bigl(\Pi
P_-\phi,RP_-\psi\bigr) \\ 
&= -\bigl(P_-\phi,(H+1)^{-1}[H,i\Pi](H+1)^{-1}P_-\psi\bigr)
\end{split}
\end{equation}
Taking \eqref{eq:bounded} into account, we see that it is legitimate to write
\begin{equation}
[\tR,i\tPi]= -\tR[H,i\Pi]\tR
\label{eq:commRP}
\end{equation}
as a bounded operator on $\Hi$. Because $\tR (H_0+1)=P_-$ is bounded, by
interpolation and duality, $\tR^{\epsilon/2} (H_0+1)^{\epsilon/2}$ and 
$(H_0+1)^{\epsilon/2} \tR^{\epsilon/2}$ are bounded for every $0\le \epsilon\le
2$. Since $\abs{\tilde p_y}^{\epsilon} (H_0+1)^{-\epsilon/2}$ is also bounded
and by again using \eqref{eq:bounded}, we conclude that the expressions
\begin{equation}
\abs{\tilde p_y}^{\epsilon}\tR[H,i\Pi]\tR\abs{\tilde p_y}^{\epsilon}, \quad
\abs{\tilde p_y}^{\epsilon}\tR[H,i\Pi]\tR, \quad
\tR[H,i\Pi]\tR\abs{\tilde p_y}^{\epsilon}
\label{eq:bounded2}
\end{equation}
are bounded operators on $\Hi$, for every $0\le\epsilon<1/2$.

\item Since the spectra of $H$ and $R$ are related by
$\sigma(R)=(\sigma(H)+1)^{-1}$ with corresponding continuity properties,
and since $\sigma(\tR)=\sigma(R)\cup\{0\}$, we see that the spectrum of $H$ is
absolutely continuous on a bounded interval $\Delta>0$ 
if and only if the spectrum of $\tR$ is
absolutely continuous on $(\Delta+1)^{-1}$. 
On the other hand, the Mourre estimate
for $H$ on a bounded interval $\Delta$ is equivalent to a Mourre estimate for
$\tR$ on the interval $(\Delta+1)^{-1}$ because of \eqref{eq:commRP} and 
because $\tR$ is positive on a bounded interval. The sign of the commutator
changes when we pass from $H$ to $\tR$, but this does not affect the
applicability of Mourre theory.

\item By the preceding remarks, it is enough to show that $\tR$ is of class
  $\Ce^{1+\epsilon}(\tPi)$ for some $\epsilon>0$ to complete the proof of
  absolute continuity of the spectrum of $H$ 
  in the intervals on which we have established a Mourre estimate by Theorem
  \ref{thm:mourreest2}.
\begin{lemma}
There is a constant $C$ and an $\epsilon>0$ such that
\begin{equation}
\norm{e^{i\tPi\tau}\tR[H,i\Pi]\tR e^{-i\tPi\tau} - \tR[H,i\Pi]\tR} \le
C\abs{\tau}^{\epsilon} 
\end{equation}
as $\abs{\tau}\rightarrow 0$. In other words, $\tR\in\Ce^{1+\epsilon}(\tPi)$.
\end{lemma}
\begin{proof}[Proof]
First note that $e^{i\tPi\tau}=e^{i\tilde p_y\tau}e^{ix\tau}$, and therefore
\begin{equation}
\begin{split}
\norm{e^{i\tPi\tau}\tR[H,i\Pi]\tR e^{-i\tPi\tau} - \tR[H,i\Pi]\tR} \le &
\norm{e^{i\tilde p_y\tau}\tR[H,i\Pi]\tR e^{-i\tilde p_y\tau} - 
\tR[H,i\Pi]\tR} \\
+& \norm{e^{ix\tau}\tR[H,i\Pi]\tR e^{-ix\tau} - 
\tR[H,i\Pi]\tR}.
\end{split}
\label{eq:2terms}
\end{equation}
For the first term, we have used that $e^{ix\tau}$ is unitary for every
$\tau$. Write
\begin{equation}
\begin{split}
e^{i\tilde p_y\tau}\tR[H,i\Pi]\tR e^{-i\tilde p_y\tau} - 
\tR[H,i\Pi]\tR = &
\left(e^{i\tilde p_y\tau}-1\right)\tR[H,i\Pi]\tR\left(e^{-i\tilde p_y\tau}
  -1\right) \\ 
&+ \left(e^{i\tilde p_y\tau}-1\right)\tR[H,i\Pi]\tR \\
&+ \tR[H,i\Pi]\tR \left(e^{-i\tilde p_y\tau}-1\right)
\end{split}
\end{equation}
Using the spectral theorem, one shows that for all $1>\epsilon>0$
there is a constant $C$ such that  
\begin{equation}
\abs{e^{i\tilde p_y\tau}-1} \le C \abs{\tilde
  p_y}^{\epsilon}\abs{\tau}^{\epsilon}. 
\end{equation}
In combination with \eqref{eq:bounded2}, the last two lines show that
\begin{equation}
\norm{e^{i\tilde p_y\tau}\tR[H,i\Pi]\tR e^{-i\tilde p_y\tau} - 
\tR[H,i\Pi]\tR} \le C \abs{\tau}^{\epsilon} 
\end{equation}
for some $\epsilon$ and $C$. As for the other term in equation
\eqref{eq:2terms}, one shows
that $\tau\mapsto e^{ix\tau}\tR[H,i\Pi]\tR e^{-ix\tau}$ is even differentiable
at $\tau=0$ because the commutator of $x$ with $H$ is simply proportional to
$p_x+y$ and bounded by $H$. This remark completes the proof.
\end{proof}
\end{enumerate} 



\newpage

\appendix
\section{Corbino disc geometry}
\label{sec:app1}
\begin{figure}[ht]
  \begin{center}  
    \psfrag{B}{$\vec{B}$}
    \psfrag{Phi}{$\Phi$}
    \psfrag{Er}{$\vec{E}_r$}
    \psfrag{Vr}{$V_H$}
    \psfrag{jphi}{${I}_{\varphi}$}
    \includegraphics{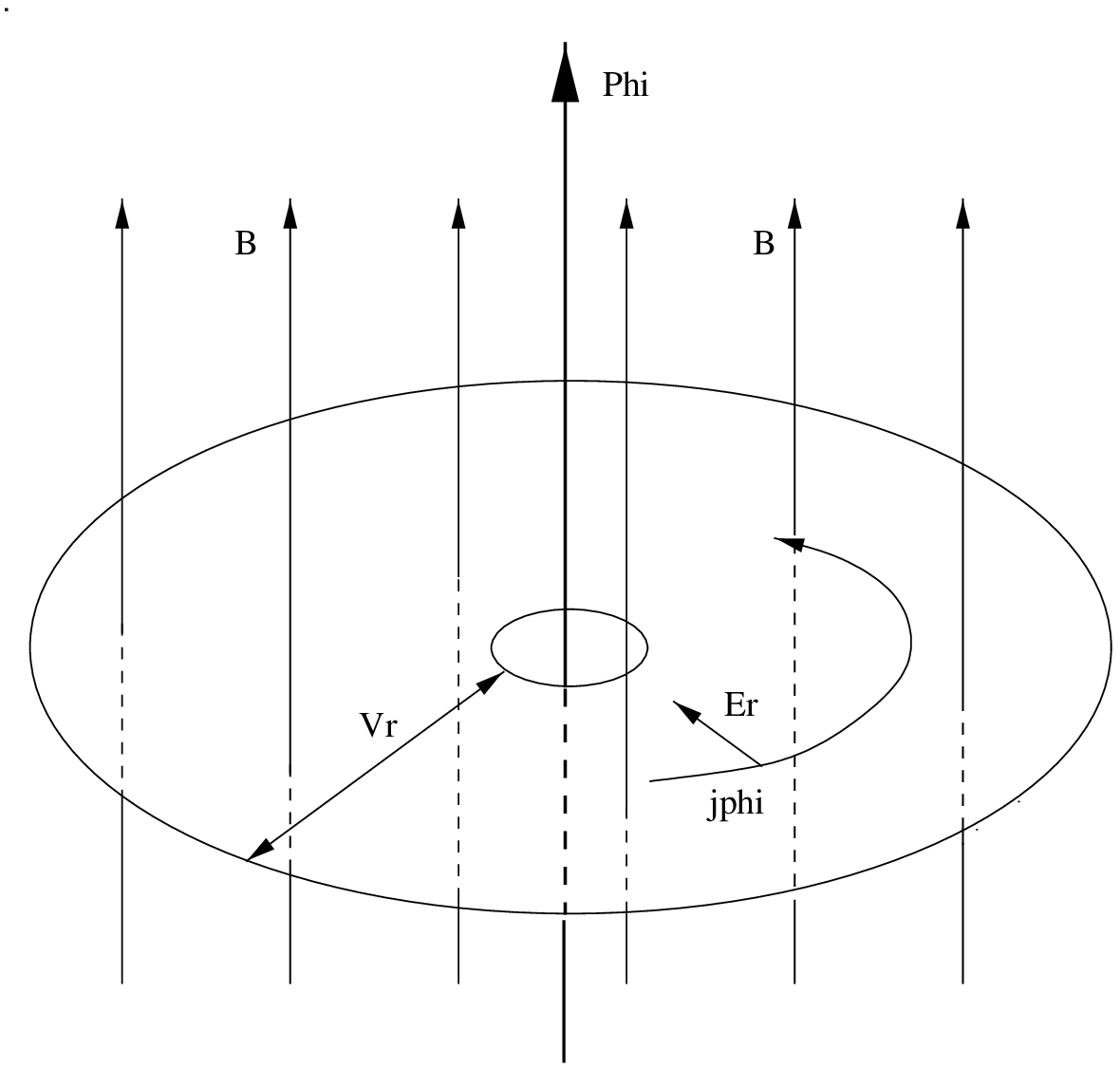}
    \caption{Corbino disc geometry}
    \label{fig:corbino}
  \end{center}
\end{figure}

In this appendix, we adapt the arguments of section \ref{sec:big} to the
Corbino disc geometry (see figure \ref{fig:corbino}). 

In polar coordinates and with a suitable gauge, the Hamiltonian is
\begin{equation}
H= \frac{1}{2m}\left(-\partial_r^2 - \frac{1}{r}\partial_r +
  \left(\frac{1}{ir}\partial_\varphi   
-e\left(\frac{Br}{2}+\frac{\Phi}{2\pi r}\right)\right)^2\right) +
V(r,\varphi).
\end{equation}

For $V=0$, the spectrum and eigenfunctions of $H$ can be obtained by
elementary methods. For $\Phi=0$, the spectrum consists only of the Landau
levels, with energy $(n+1/2)\omega_c$. In contrast to the case of the
cylinder, there is here a restriction 
on the angular momentum, $l\ge -n$. The states are localized in radial
direction near $r_0(l) = \sqrt{2\abs{l}/eB}$. If $\Phi\neq 0$, $0<\Phi\le
2\pi/e$, the localization of the states is shifted from $r_0(l)$ to
$r_0(l-e\Phi/2\pi)$. The energy of the states with $l-e\Phi/2\pi\ge 0$ remains
unchanged, but the energy of the states with $l-e\Phi/2\pi<0$ is changed to
$(n+e\Phi/2\pi+1/2)\omega_c$. It is convenient to change the definition of the
index $n$ and to introduce functions $E_{n,l}(\Phi)$ in such a way that the
spectral flow can be written as in equation \eqref{eq:flow},
\begin{equation}
E_{n,l}(\Phi+2\pi/e)=E_{n,l-1}(\Phi).
\label{eq:cflow}
\end{equation}
The energies are $E_{n,l}(\Phi) = (n+1/2)\omega_c$ if $l-e\Phi/2\pi\ge 0$ and 
$E_{n,l}(\Phi)= (n-l+e\Phi/2\pi+1/2)\omega_c$ if $l-e\Phi/2\pi< 0$.
With this definition of $n$, $l$ is unrestricted, but the bands
for fixed $n$ are bent upwards when $l-e\Phi/2\pi$ is negative. Thus, the flux
$\Phi$ has effects similar to an edge at the center of the disc. Besides
considering only one edge, we also need to restrict the analysis to $\Phi=0$
in order to avoid resonances.

The discussion with included edge potential, $V_0$, and weak disorder
potential, $gV_d$, is completely parallel
to the case of the cylinder geometry and we do not repeat it here. 

We now derive the expression for the azimuthal current corresponding to
equation \eqref{eq:super} of section \ref{sec:big}. 
In the Corbino disc geometry, the azimuthal current carried by a state,
$\psi$, is 
\begin{equation}
\begin{split}
I_{\varphi} &=-\frac{\partial E}{\partial \Phi}= \left(\psi,\frac{2}{2\pi
    r}\Bigl(  
\frac{\partial_{\varphi}}{ir} - \frac{\Phi}{2\pi r} - \frac{Br}{2}\Bigr) \psi
\right) \\
&= \bigl(\psi,\frac{\hat B\times\vec{r}}{2\pi r^2} [H,i\vec{r}]\psi\bigr).
\end{split}
\end{equation}
Note here that $\hat B\times \vec{r}/r$ is a unit vector pointing in the
azimuthal direction, and that the commutator $[H,i\vec{r}]$ gives the
velocity ($m=1/2$ and $e=1$). Now replace $\vec{r}$ in the commutator with the
guiding center, $\vec{r}= \vec{Z} - (\vec{p}-\vec{A})
\times \hat B/B$, and use the equation of motion for $\vec{Z}$,
$[H,i\vec Z]=  \hat B/B \times \grad V$. This yields
\begin{equation}
I_{\varphi} = \bigl(\psi,\frac{1}{2\pi B r} \partial_r V \psi\bigr) 
- \bigl(\psi, \frac{\hat B\times\vec{r}}{2 \pi B r^2}
[H,i(\vec{p}-\vec{A})\times \hat B] \psi \bigr). 
\label{eq:spitze}
\end{equation}
Using the fact that the expectation value of a commutator with $H$ in an
energy eigenstate vanishes, the second term in \eqref{eq:spitze} is equal to
\begin{equation}
\bigl(\psi, \frac{1}{2\pi B}\bigl[H,i\frac{\hat B\times \vec{r}}{r^2}\bigr]
(\vec{p}-\vec{A})\times \hat B \psi\bigr)
\label{eq:toll}
\end{equation}
A straightforward calculation in polar coordinates then shows that the
azimuthal current carried by an eigenstate of $H$ can be written as
\begin{equation}
\begin{split}
I_{\varphi} & = \bigl(\psi,\frac{2}{2\pi r}
(\vec{p}-\vec{A})_{\varphi}\psi\bigr) \\ 
&=  \bigl(\psi,\frac{1}{2\pi B r} \partial_r V \psi\bigr) 
+\frac{2}{2\pi B}\bigl(\frac{1}{r}\partial_r \psi,\frac{1}{r}\partial_r
\psi\bigr) 
- \frac{1}{2\pi B} \bigl(\psi,\frac{2}{r^2} (\vec{p}-\vec{A})_{\varphi}^2\psi
\bigr).
\end{split}
\label{eq:auchtoll}
\end{equation}
This expression is similar to \eqref{eq:super}, with correction terms due to
the circular geometry. The first term in \eqref{eq:auchtoll} is positive for
weak disorder, and decays inversely proportional to the size $R$ of the
sample. The second term is always positive, while the third term is
negative, but bounded, and decays as $1/R^2$. Thus
$\eval{I_{\varphi}}_{\Phi=0}$ is positive for an eigenstate of $H$ with energy
between Landau levels, if the sample is large, and the disorder weak. 

The rest of this appendix is devoted to making the estimates in this argument
rigorous. To deal with the singularity at $r=0$, it will be useful to know that
edge states eigenfunctions are exponentially small near the origin. We start
by considering the spatial decay of the free resolvent $(E-H_0)^{-1}$ (for
$\Phi=0$). This free resolvent for a homogeneous magnetic field Hamiltonian 
in two dimensions can be calculated using an expansion in eigenfunctions of
$H_0$, which are known explicitly in terms of Laguerre polynomials. See
\cite{JP} and \cite{Lehmann} for more details. The result contains the
confluent hypergeometric function called $\Psi$ in the notation of
\cite{EMOT},
\begin{equation}
\begin{split}
R_0(x,y;E)& =
(E-H_0)^{-1}(r_1e^{i\varphi_1},r_2e^{i\varphi_2}) \\ & =
-\frac{1}{4\pi} e^{i\pi\zeta}\Gamma(-\zeta)e^{ir_1r_2\sin(\varphi_1 -
  \varphi_2)} e^{-\frac{B}{4} \abs{x-y}^2}
  \Psi\bigl(-\zeta,1;\frac{B}{4}\abs{x-y}^2\bigr).
\end{split}
\end{equation}
Here, $\zeta$ is related to the energy by $E=(2\zeta+1)B$, so that $R_0$ has 
singularities at Landau band energies. $\Psi(-\zeta,1;z)$ has a
logarithmic singularity at $z=0$ and behaves for large $\abs{z}$ as
$\abs{z}^{\zeta}$. This implies a Gaussian decay for large $\abs{x-y}$. We
will, however, only use an estimate of the form
\begin{equation}
\label{eq:R0}
\abs{R_0(x,y;E)} \leq C e^{-
  \abs{x-y}/\xi} \abs{\ln(\abs{x-y}/\xi)},
\end{equation}
with a decay length scale $\xi$ on the order of the magnetic length
$\sqrt{B}^{-1}$. 

We now include the disorder potential and claim the following 

\begin{lemma}
\label{thm:decay}
Let $V_0$ be a not too fast increasing 
edge potential that describes the Corbino disc with varying
sample size $R$. Let further $V_d$ be bounded by a constant sufficiently small
compared to the magnetic field. Let finally $\Delta$ be an energy interval in
the spectral gaps of $H_0+V_d$. Whenever $a<R$, there are positive constants
$C$ and $\lambda$, such that an eigenfunction $\psi$ of $H=H_0+V_0+V_d$ with
energy $E\in\Delta$ satisfies 
\begin{equation}
\abs{\psi(x)} \le C e^{-(R-a)/\lambda} \norm{\psi}
\end{equation}
for $\abs{x}\le a$, uniformly in the sample size $R$ and the energy
$E\in\Delta$.  
\end{lemma}

\begin{proof}[Proof]
Use the equation
\begin{equation}
\psi(x)= \int(E-H_0-V_d) ^{-1}(x,y) V_0(y)\psi(y) dy
\end{equation}
and expand the resolvent in a Neumann-series.
\begin{equation}
\label{eq:neumann}
(E-H_0-V_d)^{-1} =  \sum_{n=0}^{\infty} \left((E-H_0)^{-1}
V_d \right)^{n}(E-H_0)^{-1}.
\end{equation}
Consider a fixed $n$, and use the estimate \eqref{eq:R0} for each free
resolvent. This results in the following integrals to be estimated:
\begin{multline}
\int \Bigl(\prod_{i=1}^{n} C \abs{\ln(\abs{z_{i-1}-z_i}/\xi)}
e^{-\abs{z_{i-1}-z_i}/\xi}V_d(z_i)\Bigr) \times \\
C \abs{\ln(\abs{z_n-y}/\xi)} e^{-\abs{z_n-y}/\xi}
V_0(y)\abs{\psi(y)}  dz_1 \dots dz_n dy, 
\end{multline}
with $z_0=x$, $z_{n+1}=y$. Taking out $V_d$ out of the integrals, and
splitting the exponentials in $3$, this is estimated by
\begin{multline}
(C\norm{V_d}_{\infty})^n \sup_{z_1,\dots z_{n+1}} \exp(-\sum_{i=1}^{n+1}
\abs{z_{i-1}-z_i}/3\xi) \times \\
\Bigl(\int e^{-\abs{w}/3\xi}\abs{\ln(\abs{w}/\xi)}\Bigr)^n 
\sup_{z_1,\dots z_{n}} \int e^{-\abs{z_n-y}/3\xi}
\abs{\ln(\abs{z_n-y}/\xi)} \times \\  \exp(-\sum_{i=1}^{n+1}
\abs{z_{i-1}-z_i}/3\xi) V_0(y)\abs{\psi(y)} dy.
\end{multline}
Now $\inf_{z_1,\dots z_n}\sum_{i=1}^{n+1} \abs{z_{i-1}-z_i} \ge \abs{x-y}$
and, applying the Schwarz inequality to the last integral, the bound moves to
\begin{multline}
 (\tilde{C}\norm{V_d}_{\infty})^n e^{-\abs{x-R}/3\xi} 
\Bigl(\int e^{-2\abs{w}/3\xi} \abs{\ln(\abs{w}/\xi)}^2 dw\Bigr)^{1/2}
\times \\
\Bigl(\int e^{-2\abs{x-y}/3\xi} (V_0(y))^2 \abs{\psi(y)}^2dy \Bigr)^{1/2}.
\end{multline}
If the potential does not increase too fast, the integral containing $V_0$
converges and is bounded by $E\norm{\psi}$, and after 
summing over $n$, making the bound on the disorder small enough, the claim
follows. 
\end{proof}

With the help of Lemma \ref{thm:decay}, we now prove the positivity of the
current from the expression \eqref{eq:auchtoll},
\begin{equation}
\begin{split}
I_{\varphi} & = \bigl(\psi,\frac{2}{2\pi r}
(\vec{p}-\vec{A})_{\varphi}\psi\bigr) \\ 
&=  \bigl(\psi,\frac{1}{2\pi B r} \partial_r V \psi\bigr) 
+\frac{2}{2\pi B}\bigl(\frac{1}{r}\partial_r \psi,\frac{1}{r}\partial_r
\psi\bigr) 
- \frac{1}{2\pi B} \bigl(\psi,\frac{2}{r^2} (\vec{p}-\vec{A})_{\varphi}^2\psi
\bigr).
\end{split}
\end{equation}
The first step is to eliminate the singularity at $r=0$ by replacing $\psi$
with $j\psi$, where $j$ is a cutoff at radius $a>0$ near the origin.
We want to show that the error introduced by this replacement,
\begin{multline}
\label{eq:error}
 \bigl(\psi,\frac{1}{r} (p-A)_{\varphi}\psi\bigr) - 
 \bigl(j\psi,\frac{1}{r} (p-A)_{\varphi}j\psi\bigr) = \\ 
 \bigl((1-j)\psi,\frac{1}{r} (p-A)_{\varphi}j\psi\bigr) +
 \bigl(j\psi,\frac{1}{r} (p-A)_{\varphi}(1-j)\psi\bigr) + \\
 \bigl((1-j)\psi,\frac{1}{r} (p-A)_{\varphi}(1-j)\psi\bigr),
\end{multline}
is small for large $R$. Consider bounding the term 
\begin{equation}
\begin{split}
 \abs{\bigl((1-j)\psi,\frac{1}{r} (p-A)_{\varphi}(1-j)\psi\bigr)} & \le
\norm{\smash{\frac{1}{r^{3/4}} \psi}}_a
\norm{\smash{\frac{1}{r^{1/2}}(p-A)_{\varphi}\psi}}_a \\
& \le C e^{-R/\lambda} a^{1/4} \norm{\psi}
\norm{\smash{\frac{1}{r}\psi}}^{1/2}_a 
\norm{(p-A)^2_{\varphi}\psi}_a^{1/2} \\
\le C e^{-3R/2\lambda} a^{1/2} \norm{\psi}^2
\end{split}
\end{equation}
where $\norm{\,\cdot\,}_a$ has its obvious meaning, and 
we have taken advantage of the fact the $r$ commutes with
$(p-A)_{\varphi}$ and that $\norm{\smash{(\vec{p}-\vec{A})^2_{\varphi}\psi}}$ 
can be bounded by the energy\footnote{This Kato estimate for
  $(\vec{p}-\vec{A})^2_{\varphi}$ is not trivial, since radial and azimuthal
  part of the kinetic energy do not commute. The same kind of estimate in the
  half-plane geometry was needed in section \ref{sec:diric}.}.
The other terms in \eqref{eq:error} similarly decay exponentially with the
sample size.  

The replacement of $\psi$ with $j\psi$ introduces additional terms, because in
the derivation of \eqref{eq:auchtoll}, we used that $\psi$ was an eigenstate
of $H$, whereas $j\psi$ is not. Those additional terms are
\begin{multline}
\bigl(j\psi, \frac{1}{2\pi B}i\frac{\hat B\times \vec{r}}{r^2}
(\vec{p}-\vec{A})\times \hat B [H,j]\psi\bigr) 
+ \bigl(\psi,[H,j]  \frac{1}{2\pi B}i\frac{\hat B\times \vec{r}}{r^2}
(\vec{p}-\vec{A})\times \hat B j\psi\bigr) \\ =
\bigl(j\psi, \frac{1}{2\pi B r} (-\partial_r) [H,j] \psi \bigr) +
\bigl(\psi, [H,j] \frac{1}{2\pi B r} (-\partial_r) j\psi\bigr)
\end{multline}
$[H,j] = -2j'\partial_r - j'' -j'/r$ has support for $r$ near $a$.
Organize the various terms so that either no radial derivative
is acting on $\psi$, or one radial derivative, or the expression
$\frac{1}{r}\partial_r r\partial_r$. Terms with one radial derivative are
estimated, for example, as
\begin{equation}
\begin{split}
\abs{\bigl(j\psi, j''\frac{1}{r} \partial_r j\psi\bigr)} & \le 
\norm{\smash{j''\frac{1}{r}j'\psi}}_a \norm{\partial_r \psi} \\
&\le \frac{C}{a^4} e^{-(R-a)/\lambda} \norm{\psi}^2,
\end{split}
\end{equation}
where $C$ contains the energy as bound for $\norm{\partial_r \psi}$.
The terms with two radial derivatives, such as
\begin{equation}
\bigl(j'j\frac{1}{r}\psi, \frac{1}{r} \partial_r r\partial_r \psi \bigr),
\end{equation}
are estimated similarly. Terms without radial derivative acting on $\psi$
are even easier.

Cutting off with $j$ thus introduces error terms that decay exponentially
with the sample size. 

From \eqref{eq:auchtoll} with $j\psi$ instead of $\psi$, we now estimate
\begin{equation}
\abs{\bigl(j\psi,\frac{1}{r^2} (\vec{p}-\vec{A})_{\varphi}^2j\psi
\bigr)} \le \norm{\smash{\frac{1}{r^2}j\psi}} \norm{j
(\vec{p}-\vec{A})_{\varphi}^2\psi}.
\end{equation}
The second factor is bounded by the energy. The first is split into a
part with the radial coordinate between $a$ and $R/2$, so that it decays as
$e^{-R/2\lambda}/a^2$, and a part from $R/2$ to $\infty$ which decays as
$1/R^2$. The positivity of 
\begin{equation}
\bigl(j\psi,\frac{1}{ r} \partial_r V j\psi\bigr) 
\end{equation}
is proved in a fashion very similar to the one in section \ref{sec:edge} by
introducing another cutoff function at the edge $r=R$\footnote{The cutoff at
  $r=a$ does not 
  significantly perturb the argument of section \ref{sec:edge} because $\psi$
  is exponentially small near the center.}.
The lower bound for the expression decays as $1/R$. If the sample is large
and the disorder weak, the current is positive.


\section{Random potentials and almost sure spectrum}
\label{sec:app2}
Consider a Hamiltonian $H_1$ on $\Hi=L_2(\R^2)$ which is invariant in
$x$-direction. Let $\sigma(H_1)$ be the spectrum of $H_1$. 
Add to $H_1$ a disorder in the form of a random potential $V_{d,\omega}$,
where $\omega\in\Omega$, and $(\Omega,P)$ is a probability space.
Assume three things about $(\Omega,P)$:
\begin{enumerate}
\setlength{\parskip}{0mm}
\renewcommand{\labelenumi}{(\roman{enumi})}
\renewcommand{\theenumi}{(\roman{enumi})}
\item \label{assum1} For every $\omega\in\Omega$, $V_{d,\omega}$ is bounded by
  a constant $\delta$ which is independent of $\omega$. 
\item \label{assum2} The group $G^{(x)}$ of translations in $x$-direction acts
 measure-preserving and ergodically on $(\Omega,P)$. This allows one to
 speak of an almost sure (a.s.) spectrum of $H_{\omega}=H_1+V_{d,\omega}$,
 denoted by $\Sigma(H_{\omega})$.
\item \label{assum3} For every measurable compact set $\Lambda\subset\R^2$ and
  every $\epsilon>0$ the probability
\begin{equation}
P\bigl\{\omega; \,\abs{V_{d,\omega}(x,y)}<\epsilon\;\;
\forall (x,y)\in \Lambda\bigr\}
\end{equation}
is positive.
\end{enumerate}
Those assumptions are, for example, satisfied in an Anderson model for the
disorder. Assumption \ref{assum1} is added for consistency with the proofs in
sections \ref{sec:edge} and \ref{sec:diric}. 
As mentioned in section \ref{sec:methods}, it can
very likely be replaced by boundedness of the variance of $V_{d,\omega}$. 
Assumptions \ref{assum2} and \ref{assum3} allow the proof of the
following:
\begin{lemma}
Let notation be as introduced and assumptions \ref{assum1} to \ref{assum3} be
satisfied. Then
\begin{equation}
\sigma(H_1)\subset \Sigma(H_{\omega}) \subset \sigma(H_1)+
\inter{-\delta}{\delta}
\end{equation}
\end{lemma}
\begin{proof}[Proof]
For the first inclusion let $E\in\sigma(H_1)$ and $\epsilon>0$. Then there is a
$\psi\in\Hi$ with $\norm{(E-H_1)\psi}<\epsilon/3\,\norm{\psi}$. 
Let $\Lambda$ be a large measurable compact subset of $\R^2$.
With $\norm{\,\cdot\,}_{\Lambda} = \norm{\chi_{\Lambda}\,\cdot\,}$, we have:
\begin{equation}
\begin{split}
\norm{(E-H_{\omega})\psi} &\le \norm{(E-H_1)\psi}+ \norm{V_{d,\omega}\psi} \\
&\le \frac{\epsilon}{3}\norm{\psi} + \norm{V_{d,\omega}\psi}_{\Lambda}+
\delta\norm{\psi}_{\R^2\setminus\Lambda} \\
\end{split}
\end{equation} 
Take $\Lambda$ so large that
$\delta\norm{\psi}_{\R^2\setminus\Lambda}<\epsilon/3 \,\norm{\psi}$, and
define
\begin{equation}
A:=\{\omega; \abs{V_{d,\omega}(x,y)}<\epsilon/3 \;\;\forall
(x,y)\in \Lambda\}.
\end{equation}
For $\omega\in A$ we then have
\begin{equation}
\norm{(E-H_{\omega})\psi} \le \frac{2\epsilon}{3}\norm{\psi} +
\frac{\epsilon}{3} \norm{\psi}_{\Lambda} \le \epsilon \norm{\psi}, 
\end{equation}  
and the probability of this event, $P(A)$, is positive because of assumption
\ref{assum3}, so that
\begin{equation}
P\bigl\{\omega; \, \dist\bigl(E,\Sigma(H_\omega)\bigr)<\epsilon\bigr\}>0.
\end{equation}
The last event is translation invariant and must then have probability one
because of ergodicity. Therefore $E\in\Sigma(H_{\omega})$. 

For the proof of the second inclusion let 
$E\notin\sigma(H_1)+\inter{-\delta}{\delta}$, that is $\dist(E,\sigma(H_1))
=\eta>\delta$. For all $\psi$ in the domain of $H_1$,
$\De(H_1)=\De(H_{\omega})$, $\norm{(E-H_1)\psi}\ge\eta\norm{\psi}$ holds, and
therefore 
\begin{equation}
\begin{split}
\norm{(E-H_{\omega})\psi} &\ge \norm{(E-H_1)\psi} - \delta\norm{\psi} \\
&\ge (\eta-\delta)\norm{\psi} > 0, 
\end{split}
\end{equation}
so that $E\notin\Sigma(H_{\omega})$.
\end{proof}

\section{Alternative proof of proposition \ref{thm:proposition}}
\label{sec:app3}
Let $\Delta\ni E$ and $\psi=E_{\Delta}(H)\psi$. 
As in section \ref{sec:edge} let  
$\eta=\dist (E,\sigma(H_0))$, and $'=\partial_y$.
Take $j$ to be a smooth cutoff function with $j(y)=1$ for $y\le-1$ 
and $j(y)=0$ for $y\ge 0$. Let $c_1=\sup (\abs{j'})$ and 
$c_2= \sup(\abs{j''})$. For $a>0$ define $j_a$ as cut-off function on  
$\left[-a,0\right]$ by $j_a(y)=j(y/a)$. Then $\abs{j_a'}\le c_1/a$
and $\abs{j_a''}\le c_2/a^2$.
Also write 
$\nnorm{\,\cdot\,}_a = \nnorm{\chi_{\left[-a,0\right]}\,\cdot\,}$ for $a>0$ and 
$\nnorm{\,\cdot\,}_0^2 = \int\!\!\de{x}\,\nabs{\,\cdot\,}^2_{y=0}$.

\newlength{\parindentha}
\setlength{\parindentha}{\parindent}
Consider now $j_a\psi$. Obviously
$j_a\psi\in {\cal D}(H_0)$ and 
\begin{equation}
\begin{split}
\eta\norm{j_a\psi} \le
\norm{(E-H_0)j_a\psi}& =\norm{(E-H_0)j_a\psi}_- \\
& \le\norm{(E-H)j_a\psi}_- + \delta\norm{\psi}_- \\
&\le \underbrace{\norm{j_a(E-H)\psi}_-}_{\displaystyle 
\le \abs{\Delta}\norm{\psi}_-} + 
\norm{[H,j_a]\psi}_- + \delta\norm{\psi}_-.
\label{eq:grund}
\end{split}
\end{equation}
Because $[H,j_a] = -2i{j_a'}p_y - j_a''$,
\begin{equation}
\begin{split}
\norm{[H,j_a]\psi}_- &\le 2\norm{j_a'\psi'}_- + \norm{j_a''\psi}_- \\
& \le \frac{2c_1}{a} \norm{\psi'}_a + \frac{c_2}{a^2} \norm{\psi}_a.
\end{split}
\label{eq:hilf}
\end{equation}

For $\psi=E_{\Delta}(H)\psi$, $\psi(x,0)=0$ and for 
$\Delta$ a bounded intervall, we also have 
$\phi:=H\psi =E_{\Delta}\phi \in{\cal D}(H)$. This implies 
\begin{equation}
0=\phi(x,0)=\bigl(-\psi''+ (p_x+By)^2\psi+ V_d \psi \bigr)(x,0)= -\psi''(x,0).
\end{equation}
Because $\psi=E_{\Delta}\psi$, $\psi$ is sufficiently smooth and the formula 
$\psi(x,y)= \int_0^y\psi'(x,\upsilon)\,\de{\upsilon}$ holds. Therefore
\begin{align}
\begin{split}
\norm{\psi}^2_a &= \int_{y\ge -a} \biggl\lvert \int_0^y
  \psi'(x,\upsilon)\de{\upsilon} \biggr\rvert^2  \de{x}\de{y} \\
&\leq \int_{y\ge -a} \int_0^y\abs{\psi'(x,\upsilon)}^2\de{\upsilon} y
  \de{x}\de{y} \\
&\leq \frac{1}{2} a^2 \norm{\psi'}^2_a.
\end{split}
\intertext{%
Because $\psi'(x,y)= \psi'(x,0)+ \int_0^y \psi''(x,y)\de{y}$ we analogously have} 
\norm{\psi'- {\psi'}\rvert_{y=0}}_a^2 &\le \frac{1}{2} a^2 \norm{\psi''}^2_a,\\
\intertext{so that}
\begin{split}
\norm{\psi'}_a &\le \sqrt{a} \norm{\psi'}_0 + 
\frac{a}{\sqrt{2}}\norm{\psi''}_a \\ 
&= \sqrt{a\Gamma_{\psi}} + \frac{a}{\sqrt{2}} \norm{\psi''}_a,
\end{split}
\intertext{as well as}
\norm{\psi''}^2_a &\le \frac{1}{2}a^2\norm{\psi'''}^2_a. \\
\intertext{With a Kato estimate of the form $\norm{\psi'''}^2_- \le
  C\norm{\psi}_-^2$, this implies} 
\begin{split}
\norm{\psi'}_a &\le \sqrt{a\Gamma_{\psi}} + \frac{a^2}{2}\norm{\psi'''} \\
               &\le \sqrt{a\Gamma_{\psi}} +
               \frac{a^2}{2}C \norm{\psi}_-,
\end{split} \\
\intertext{whereas $\norm{\psi}_a \le a/\sqrt{2}\norm{\psi'}_a$ implies}
\norm{\psi}_- &\le \norm{j_a\psi} +\norm{\psi}_a 
\le \norm{j_a\psi} + \frac{a}{\sqrt{2}}\norm{\psi'}_a.
\end{align}
Continuing the estimates \eqref{eq:grund} and \eqref{eq:hilf}, one obtains:
\begin{align}
\begin{split}
\eta\bigl(\norm{\psi}_- -\frac{a}{\sqrt{2}}\norm{\psi'}_a\bigr) & 
\leq \eta\norm{j_a\psi} \\
& \le \bigl(\abs{\Delta}+\delta\bigr)\norm{\psi}_- + \Bigl(\frac{2c_1}{a} +
\frac{c_2}{\sqrt{2}a}\Bigr)  \norm{\psi'}_a
\end{split} \\[0.2cm]
\implies
\bigl(\eta - \abs{\Delta}-\delta\bigr)\norm{\psi}_- 
& \leq \Bigl(2c_1+\frac{c_2}{\sqrt{2}}+
\frac{\eta a^2}{\sqrt{2}} \Bigr)
\Bigl(\sqrt{\frac{\Gamma_{\psi}}{a}} + \frac{a}{2}C
\norm{\psi}_- \Bigr)
\label{eq:alpha}
\end{align}
Now let $\eta>\delta$.
If we assume the contrary of our claim, i.e. that $\Gamma_{\psi_n}
\rightarrow 0$ for a sequence of $\psi_n$'s in $E_{\Delta}(\Hi^-)$,
we obtain a contradiction for $\abs{\Delta}$ and $a$ small enough.
\qed

\end{document}